\documentclass[useAMS,usenatbib,dvipdfmx]{mn2e}
\usepackage{setspace, ulem,amsmath,amssymb}
\usepackage[dvipdfmx]{graphicx,color}
\begin{document}
%\title[]{Transonic galactic outflow in a dark matter halo and a central black hole gravitational potential}
\title[Polytropic transonic outflows in a dark matter halo with a central black hole]
{Polytropic transonic galactic outflows in a dark matter halo with a central black hole}
\author[Igarashi, Mori and Nitta]{Asuka Igarashi$^{1}$, Masao Mori$^{1,2}$, and Shin-ya Nitta$^{3,4,5}$\\
$^{1}$Faculty of Pure and Applied Sciences, University of Tsukuba, 1-1-1, Tennodai, Tsukuba, Ibaraki, 305-8577, Japan\\
$^{2}$Center for Computational Sciences, University of Tsukuba, 1-1-1, Tennodai, Tsukuba, Ibaraki, 305-8577, Japan\\
$^{3}$Tsukuba University of Technology, 4-3-15, Amakubo, Tsukuba, Ibaraki, 305-8520, Japan\\
$^{4}$Hinode Science Project, National Astronomical Observatory of Japan, 2-21-1 Osawa, Mitaka, Tokyo, 181-8588, Japan\\
$^{5}$Institute of Space and Astronautical Science, Japan Aerospace Exploration Agency, 3-1-1 Yoshinodai, Sagamihara, \\Kanagawa, 229-8510, Japan}

\date{Accepted 201x ***. Received 2015 ***; in original form 2015 ***}

\pagerange{\pageref{firstpage}--\pageref{lastpage}} \pubyear{201x}

\maketitle

\label{firstpage}

\begin{abstract}

Polytropic transonic solutions of spherically symmetric and steady galactic winds in the gravitational potential of a dark matter halo (DMH) with a supermassive black hole (SMBH) are studied. 
The solutions are classified in terms of their topological features, and the gravitational potential of the SMBH adds a new branch to the transonic solutions generated by the gravity of the DMH. 
The topological types of the transonic solutions depend on the mass distribution, the amount of supplied energy, the polytropic index $\gamma$, and the slope $\alpha$ of the DMH mass distribution. 
When $\alpha$ becomes larger than a critical value $\alpha_\mathrm{c}$, the transonic solution types change dramatically. 
Further, our model predicts that it is possible for a slowly accelerating outflow  to exist, even in quiescent galaxies with small $\gamma$. 
This slowly accelerating outflow differs from those considered in many of  the previous studies focusing on supersonic outflows in active star-forming galaxies. 
In addition, our model indicates that outflows in active star-forming galaxies have only one transonic point in the inner region ($\sim$ 0.01 kpc). 
The locus of this transonic point does not strongly depend on $\gamma$. 
We apply the polytropic model incorporating mass flux supplied by stellar components to the Sombrero Galaxy, and conclude that it can reproduce the observed gas density and the temperature distribution well. 
This result differs significantly from the isothermal model, which requires an unrealistically large mass flux \citep{igarashi14}. 
Thus, we conclude that the polytropic model is more realistic than the isothermal model, and that the Sombrero Galaxy can have a slowly accelerating outflow.  

\end{abstract}


\section{Introduction}

In modern theoretical cosmology, the cold dark matter (CDM) scenario has been successfully applied to reproduce the observed large-scale structure of the universe; this indicates that CDM plays an important role in galaxy formation \citep{white78,davis85,frenk91,kauffmann93}.
Studies of galaxy formation based on the CDM scenario have suggested that the collapse of a CDM halo should lead to the capture of baryons by the halo's gravitational potential \citep{blumenthal84,white91,cole91}.
Further, the baryonic gases that are currently observable in such galaxies have been influenced by a variety of physical processes in the past, such as star formation, galaxy mergers, and stripping by the intergalactic medium (IGM) \citep{gunn72,dekel86,dekel87,efstathiou92}.
In particular, the studies of galaxy formation conducted to date have indicated that galactic winds have significantly influenced the history of star formation and the metal enrichment of intergalactic space \citep{larson74,faber76,mori97,mori99,mori02}.
%%2015/7/22
%In recent theoretical cosmology, the cold dark matter (CDM) senario is successful to reproduce the observed large-scale structure of the universe.
%Also, it indicates that CDM plays an important role in the formation of galaxies.
%The studies of galaxy formation based on the CDM senario suggested that the collapse of CDM halo should lead to the capture of baryons in the gravitational potential of CDM halo.
%Various physical processes, such as star formation, galaxy mergers and stripping by the intergalactic medium, would have influenced this baryonic gas that is currently observed.
%So far, studies of galaxy formations indicate that galactic winds significantly influence the history of the star formation and the metal enrichment of intergalactic space. 

The earliest works on galactic outflows were motivated by observations of the star-forming galaxy M82 \citep{lynds63,burbidge64}.
Furthermore, it is well known that the ratio of gas to stellar mass in elliptical galaxies is smaller than that in spiral galaxies \citep{osterbrock60}; this gas deficiency in elliptical galaxies indicates that galactic outflows efficiently remove the interstellar gas from these systems \citep{burke68,johnson71,mathews71}.
In addition, the presence of metals in intergalactic space also suggests that galactic outflows transport metal-containing interstellar media into these regions.
Recent observations have revealed that the low-density IGM at high redshift contains a small amount of metals \citep{songaila97,ellison00,aguirre01}.
Thus, results of these observation strongly indicate the importance of galactic outflows in galactic evolution.

However, the mechanism driving these galactic outflows is still unclear.
Theoretically, a sufficient energy supply is required for matter to escape from a galactic gravitational potential well \citep{larson74,dekel86,mori97,mori99,mori02,binney04,cattaneo06,oppenheimer06,puchwein12}, and the majority of the previous works on galactic outflows have primarily assumed that supernovae (SNe) and stellar winds function as thermal energy sources for this motion.
From a theoretical perspective, however, other possible energy sources driving galactic outflows exist. 
For example, active galactic nuclei (AGN) have been suggested as a possible energy source driving galactic outflow \citep{silk98,sharma13}.
Also, radiation pressure may act as an additional driving force, if the coupling between the dust grains and hot gas is sufficiently strong \citep{sharma12}.
This mechanism is important for galactic outflows in high-$z$ massive star-forming galaxies \citep{hopkins12}.
Cosmic rays can also drive a large-scale outflow, if the coupling between the high-energy particles and thermal gas is sufficiently strong \citep{ipavich75,breitschwerdt91,zirakashvili96,ptuskin97,uhlig12}.
However, while many driving mechanism candidates have been proposed, the majority of the previous theoretical studies have maintained that SNe function as the primary source of thermal energy in most star-forming galaxies \citep{veilleux05}.

In fact, recent observations have also indicated that a significant number of local active star-forming galaxies form starburst-driven outflows \citep{strickland02,heckman03,hessen09}.
Spectroscopic studies have shown that Lyman Break Galaxies at $z\sim3$ also exhibit galactic outflows \citep{adelberger03,shapley03}.
Furthermore, it has been found that the outflow velocity is proportional to the star formation rate and the galactic stellar mass in high-$z$ galaxies  \citep{kennicutt98a,strickland04,pettini01,shapley03,weiner09}.
These observations indicate that the bulk of the energy driving outflows is due to SNe.
%%2015/8/3
%In fact, recent observations have also indicated that a significant number of local active star-forming galaxies form starburst-driven outflows \citep{strickland02,heckman03,hessen09}.
%Spectroscopic studies have shown that Lyman Break Galaxies at $z\sim3$ also exhibit galactic outflows \citep{adelberger03,shapley03}.
%Furthermore, it has been found that the outflow velocity is proportional to the star formation rate and the galactic stellar mass in high-$z$ galaxies  \citep{kennicutt98a,strickland04,pettini01,shapley03,weiner09}.
%These observations indicate that the majority of the thermal energy driving outflows is due to SNe.
%%2015/7/22
%In fact, recent observations also indicate that a significant number of local active star-forming galaxies form starburst-driven outflows \citep{strickland02,heckman03,hessen09}. 
%In spectroscopic studies, Lyman Break Galaxies at z$\sim$3 also indicate the presence of galactic outflows \citep{adelberger03,shapley03}. 
%Furthermore, high-redshift galaxies show that the outflow velocity is proportional to the star formation rate and the stellar mass of galaxies \citep{kennicutt98a,strickland04,pettini01,shapley03,weiner09}. 
%These observations indicate that most of thermal energy driving outflows is supplied by supernovae. 

In this study, we intend to focus on transonic solutions as models of galactic outflows driven by thermal energy from SNe and stellar winds.
\citet{parker58} first examined spherically symmetric solar winds and clarified that they can pass through transonic points smoothly; this finding demonstrates that both the energy supply and gravity are essential to the transonic acceleration process.
As regards physical observation, transonic solar wind has been observed in the solar outflow by {\it Mariner $II$} \citep{neugebauer62}, and the transonic solution is well known as the entropy-maximum solution connecting the starting point (the sun) to infinity \citep{lamers99}.
When the outflow is spherically symmetric, we can prove that the transnic solution is entropy-maximum independently of the structure of galactic mass density distribution.
We show that this proof in the appendix.

Transonic solutions are also important for galactic outflows.
Owing to the complexity of the acceleration processes of transonic galactic outflows, the main stream of the theoretical studies was concentrated toward numerical studies.
However, in this paper, we focus on the fundamental features of transonic solutions using simple analytical models to obtain systemic comprehension.
For this approach, several studies have argued the existence of transonic galactic outflows by employing Parker's steady solar wind theory.
\citet{burke68} and \citet{johnson71} have applied the solar wind model to galactic outflows in the gravitational potential of the stellar halo.
\citet{chevalier85} have calculated the nature of the supersonic region in M82 without considering gravitational potential, but assuming that a transonic point is located 200 pc from the centre.
Because the dark matter halo (DMH) is the dominant component of the gravity source in galaxies, \citet{wang95} has investigated galactic outflows with DMH mass distributions including radiative cooling.
However, transonic solutions were not successfully obtained in that case, because an unrealistic single power-law DMH mass distribution ($\propto r^{-2}$) was assumed.
Thus, the validity of the theory of transonic solutions remains an open question.

Similar to \citet{chevalier85}, \citet{sharma13} have studied steady and spherically symmetric transonic galactic outflows in active star-forming galaxies while assuming a fixed transonic point at 200 pc. 
In contrast to \citet{chevalier85}, they also considered the influence of the gravitational potential of a CDM halo. 
They assumed that the thermal energy was supplied by SNe and AGN.
As a result, they concluded that SNe can drive outflows from dwarf galaxies and that AGN are important for driving high-velocity outflows in massive galaxies.
In addition, they advocated that outflows from intermediate galaxies in the quiescent star formation mode cannot escape the halo.
The transonic point was fixed in that study; however, transonic points should be determined naturally based on a balance between the thermal energy supply and the gravitational potential.
Further, \citet{tsuchiya13} discussed the influence of the DMH mass distribution gravitational potential on the nature of transonic galactic outflows while assuming steady, isothermal, and spherically symmetric conditions, without the injection of mass along the outflow lines.
They did not fix the locus of the transonic point and performed more precise analysis.

On the other hand, a consensus has not yet been reached regarding the functional form of the DMH mass distribution, and several different functions have been proposed by both simulation and observation.
For example, on the basis of the CDM scenario, \citet{navarro96} have concluded that the DMH mass density distribution has a double power-law functional form, $\rho_\mathrm{DMH}\propto r^{-1}(r+r_\mathrm{d})^{-2}$, where $r$ is the distance from the galactic centre to the DMH and $r_\mathrm{d}$ is the scale radius of the DMH.
This mass density distribution function is called the Navarro-Frenk-White (NFW) model.
Other simulations with higher resolution have also prompted the proposal of double power-law mass density distributions in the CDM scenario, although the power-law index at the centre has differed somewhat.
For example, \citet{fukushige97} and \citet{moore99} have suggested $\rho_\mathrm{DMH}\propto r^{-1.5}(r^{1.5}+r_\mathrm{d}^{1.5})^{-1}$.
These mass density distributions, which are based on numerical models, commonly diverge at the centre, in a structure known as a ``cusp".
In contrast, observations of nearby dwarf galaxies have indicated that the DMH mass distributions of these bodies have constant density at their centres; these structures are referred to as ``cores".
Hence, \citet{burkert95} has suggested an empirical profile with a core structure such that $\rho_\mathrm{DMH}\propto (r+r_\mathrm{d})^{-1}(r^2+r_\mathrm{d}^2)^{-1}$.
This unsolved problem is known as the ``cusp-core problem" \citep{moore99}.

Aiming to address this problem, \citet{tsuchiya13} adopted a mass distribution functional form of  $\rho_\mathrm{DMH}\propto r^{-\alpha}(r+r_\mathrm{d})^{-3+\alpha}$, with a concentration parameter $\alpha$ intended to express the variety of the distribution at the centre.
This profile reproduces the NFW profile with $\alpha=1$ well, and approximately reproduces the Moore profile with $\alpha=1.5$ and the Burkert profile with $\alpha=0$, as discussed in Section \ref{model}.
Based on this DMH distribution model, \citet{tsuchiya13} first reported transonic solutions incorporating the gravitational potential of the DMH.
Moreover, they showed the possibility of a new type of transonic solution in which the transonic point forms in a very distant region ($\sim100$ kpc).
This transonic solution is slowly accelerated across this wide region.
%%2015/7/22
%According to this problem, \citet{tsuchiya13} adopted a functional form of the mass distribution as $\rho_\mathrm{DMH}\propto r^{-\alpha}(r+r_\mathrm{d})^{-3+\alpha}$ with a concentration parameter $\alpha$ to express the variety of the distribution at the centre.
%This profile well reproduces the NFW profile with $\alpha=1$, approximately the Moore profile with $\alpha=1.5$ and the Burkert profile with $\alpha=0$, as discussed in Section \ref{model}. 
%Owing to this distribution model of DMH, \citet{tsuchiya13} first reported transonic solutions with the gravitational potential of DMH. 
%Moreover, they showed the possibility of a new type of transonic solution in which the transonic point forms in a very distant region ($\sim100$kpc). 
%This transonic solution is slowly accelerated across the wide region. 

It is well known that the majority of galaxies have a supermassive black hole (SMBH) at their centres \citep{marconi03}; the gravitational potential of this SMBH must influence the galactic outflow acceleration process, especially in the central region.
Therefore, \citet{igarashi14} added the gravitational potential of the central SMBH to the Tsuchiya model, because the original Tsuchiya model considers the DMH gravitational potential only.
They summarised the variety of transonic solutions under realistic mass distributions incorporating the gravitational potentials of both the DMH and SMBH, and applied their model to the Sombrero Galaxy (NGC4594) to clarify the acceleration process of the galactic outflows.
This galaxy reveals conflicting features \citep{li11}. 
Specifically, although the trace of the galactic outflow can be seen in X-ray observations, the gas density distribution in this galaxy is well reproduced as a hydrostatic state.
Further, \citet{igarashi14} have indicated that the hot gas of this galaxy may form a slowly accelerating transonic outflow.
In the widely spread subsonic region, it is difficult to distinguish the gas density from a hydrostatic state.

However, the isothermal approximation adopted by \citet{igarashi14} and \citet{tsuchiya13} is not universally applicable to galaxies.
For example, the isothermal model results in a larger amount of outflowing gas than can be supplied by SNe and stellar winds.
We expect that this discrepancy is due to the breakdown of the isothermal assumption, which causes infinite energy to be supplied to the flow.
Therefore, this indicates that a polytropic analysis with a limited specific energy is required in order to estimate the mass flux correctly.
Moreover, it is widely accepted that the temperature distribution structures of many galaxies are complex.
In fact, it has been confirmed that the observed temperature distributions of some galaxies are not isothermal-like \citep{fukuzawa06,diehl08}, i.e., a temperature gradient exists.

In this paper, we assume a polytropic, steady, and spherically symmetric state for galactic outflows. 
The transonic solutions are determined by the gravitational field, specific energy, and polytropic index.
Although our model stands upon some ideal assumptions, it does not only help quick interpretations of the observed data but also provide us with the clear and deep understanding of the fundamental nature of transonic galactic outflows. 

The structure of this paper is as follows.
We construct our model in Section \ref{model} and we summarise the results in Section \ref{results}.
In Section \ref{discussion}, the differences between the isothermal and polytropic models are discussed, along with the parameter range for actual galaxies.
The conclusions are given in Section \ref{conclusion}.
In Appendix, we show the proof that the transonic solution is entropy-maximum independently of the form of the gravitational potential.

\section[]{Analytical Model for Polytropic Wind} \label{model}

In this section, we construct the theoretical basis of our model.
As stated above, we assume a polytropic, steady, spherically symmetric outflow and ignore mass injection along the flow, except at the starting point.
We will discuss the availability of these assumptions in Sec. \ref{validity of assumptions}.
The polytropic relation is
%%2015/11/28
% In this section, we construct the theoretical basis of our model.
%As stated above, we assume a polytropic, steady, spherically symmetric outflow and ignore mass injection along the flow, except at the starting point.
%\textcolor{red}{
%We investigated the availability of these assumptions in Sec. \ref{validity of assumptions}.}
%The polytropic relation is
%%2015/10/8
%In this section, we construct the theoretical basis of our model.
%As stated above, we assume a polytropic, steady, spherically symmetric outflow and ignore mass injection along the flow, except at the starting point.
%The polytropic relation is
\begin{align}
P=K\rho^{\gamma},
\end{align}
where $P$, $\rho$, and $\gamma$ are the pressure, density, and polytropic index, respectively.
$K$ indirectly represents the magnitude of the entropy.
When $\gamma$ is specific heat ratio, the state of gas becomes adiabatic.
In the isothermal model, the all-over flow is supposed to be in perfect equilibrium with thermal reservoir.
We consider $\gamma$ as an effective parameter approximating a possible thermal interaction with the reservoir (heating and cooling).
The sound speed $c_\mathrm{s}$ is defined as
\begin{align}
c_\mathrm{s}^2 = \gamma K \rho^{\gamma-1}. \label{sound speed}
\end{align}
By differentiating Eq. (\ref{sound speed}), we obtain
\begin{align}
0 = \frac{1}{c_\mathrm{s}^2}\frac{dc_\mathrm{s}^2}{dr} - (\gamma-1) \frac{1}{\rho}\frac{d\rho}{dr}, \label{eq_polytropic_relation}
\end{align}
noting that $K=\mathrm{const.}$ along the streamline.
The basic equations are the mass conservation law and the equation of motion, which are expressed as

\begin{align}
\dot{M} &= \rho v r^2, \label{eq_mass} \\
\rho v \frac{dv}{dr} &= - \frac{dP}{dr} - \rho \frac{d\Phi}{dr}, \label{eq_motion}
\end{align}
where $v$, $\dot{M}$, and $\Phi$ are the velocity, mass flux, and gravitational potential, respectively.
Integrating Eq.\ref{eq_motion}, we obtain equation of energy
\begin{align}
E &= \frac{1}{2}v^2 + \frac{c_\mathrm{s}^2}{\gamma-1} + \Phi(x), \label{eq_energy}
\end{align}
where $E$ is specific energy.
%%2017/1/25
%\begin{align}
%\dot{M} &= \rho v r^2, \label{eq_mass} \\
%\rho v \frac{dv}{dr} &= - \frac{dP}{dr} - \rho \frac{d\Phi}{dr}, \label{eq_motion}
%\end{align}
%where $v$, $\dot{M}$, and $\Phi$ are the velocity, mass flux, and gravitational potential, respectively.
%\textcolor{red}{Integrating Eq.\ref{eq_motion}, we obtain equation of energy
%\begin{align}
%E &= \frac{1}{2}v^2 + \frac{c_\mathrm{s}^2}{\gamma-1} + \Phi(x), \label{eq_energy}
%\end{align}
%where $E$ is total energy.}
%%2016/10/14
%The basic equations are \textcolor{red}{the mass conservation law and the equation of motion}, which are expressed as
%\begin{align}
%\dot{M} &= \rho v r^2, \label{eq_mass} \\
%\rho v \frac{dv}{dr} &= - \frac{dP}{dr} - \rho \frac{d\Phi}{dr}, \label{eq_motion}
%\end{align}
%where $v$, $\dot{M}$, and $\Phi$ are the velocity, mass flux, and gravitational potential, respectively.
%\textcolor{red}{Integrating Eq.\ref{eq_motion}, we obtain equation of energy
%\begin{align}
%E &= \frac{1}{2}v^2 + \frac{c_\mathrm{s}^2}{\gamma-1} + \Phi(x), \label{eq_energy}
%\end{align}
%where $E$ is total energy.
%When $\gamma$ approaches 1 (isothermal), the limit of $E$ becomes infinity.}
%%2016/10/11
%The basic equations are those of the mass-energy conservation laws, which are expressed as
%\begin{align}
%\dot{M} &= \rho v r^2, \label{eq_mass} \\
%E &= \frac{1}{2}v^2 + \frac{c_\mathrm{s}^2}{\gamma-1} + \Phi(x), \label{eq_energy}
%\end{align}
%where $v$, $\dot{M}$, $E$, and $\Phi$ are the velocity, mass flux, total energy, and gravitational potential, respectively.
%The conservation of momentum corresponds to that of energy.
The differential equation for the Mach number is derived from Eqs. (\ref{eq_polytropic_relation})--(\ref{eq_energy}), such that
\begin{align}
& \frac{\mathcal{M}^2-1}{\mathcal{M}^2\{(\gamma-1)\mathcal{M}^2+2\}}\frac{d\mathcal{M}^2}{dr} \nonumber\\
& \qquad = \frac{2}{r} - \frac{\gamma+1}{2(\gamma-1)} \frac{1}{E-\Phi} \frac{d\Phi}{dr}, \label{eq_differentiated_mach_number}
\end{align}
where $\mathcal{M}(=v/c_\mathrm{s})$ is the Mach number.
Hence, by integrating the above equation, we obtain the Mach number equation
\begin{align}
& \mathcal{M}^{-1} \{ (\gamma-1)\mathcal{M}^2+2 \}^{\frac{\gamma+1}{2(\gamma-1)}} \nonumber\\
& \qquad = \{2(\gamma-1)\}^{\frac{\gamma+1}{2(\gamma-1)}} (\gamma K)^{-\frac{1}{\gamma-1}} \dot{M}^{-1} r^2 (E-\Phi)^{\frac{\gamma+1}{2(\gamma-1)}}. \label{eq_mach_number}
\end{align}
This equation contains $\gamma$, $K$, $\dot{M}$, $E$, and $\Phi$ as parameters.
The constant $K$ represents an integral constant in Eq. (\ref{eq_mach_number}).
Further,
\begin{align}
& c_\mathrm{s}^2 = (E-\Phi) \frac{2(\gamma-1)}{(\gamma-1)\mathcal{M}^2+2}, \\
& \rho^2 = \frac{\dot{M}^2}{(E-\Phi)\mathcal{M}^2r^4} \left( \frac{\mathcal{M}^2}{2} + \frac{1}{\gamma-1} \right).
\end{align}
In addition, the right hand side of Eq. (\ref{eq_differentiated_mach_number}) is used to identify critical points and is defined as
\begin{align}
N(r) &= \frac{4}{r} - \frac{\gamma+1}{\gamma-1} \frac{1}{E-\Phi} \frac{d\Phi}{dr}. 
\end{align}
When $\mathcal{M}=1$, the right hand side of Eq. (\ref{eq_differentiated_mach_number}) should vanish simultaneously at the loci of the critical points,
\begin{align}
N(r) =0.
\end{align}
The critical points derived by this equation show both the X-points (transonic points) and the O-points \citep{chakrabarti90}.

We divide Eqs. (\ref{eq_differentiated_mach_number}) and (\ref{eq_mach_number}) by the unit length $r_\mathrm{0}$ to obtain the non-dimensional equations
\begin{align}
& \frac{\mathcal{M}^2-1}{\mathcal{M}^2\{(\gamma-1)\mathcal{M}^2+2\}}\frac{d\mathcal{M}^2}{dx} \nonumber\\
& \qquad = \frac{2}{x} - \frac{\gamma+1}{2(\gamma-1)} \frac{1}{1-\Phi_\mathrm{n}} \frac{d\Phi_\mathrm{n}}{dx} ,
\end{align}
\begin{align}
& \mathcal{M}^{-1} \{ (\gamma-1)\mathcal{M}^2+2 \}^{\frac{\gamma+1}{2(\gamma-1)}} \nonumber\\
& \qquad\qquad\qquad = C x^2 (1-\Phi_\mathrm{n})^{\frac{\gamma+1}{2(\gamma-1)}} ,
\end{align}
\begin{align}
N(x) = \frac{4}{x} - \frac{\gamma+1}{\gamma-1} \frac{1}{1-\Phi_\mathrm{n}} \frac{d\Phi_\mathrm{n}}{dx}. \label{eq_N(x)}
\end{align}
where $x~(=r/r_\mathrm{0})$ and $\Phi_\mathrm{n}~(=\Phi/E)$ are the non-dimensional radius and gravitational potential, respectively.
The integral constant $C$ is expressed in terms of $K$, such that
\begin{align}
\log C = -\frac{\gamma}{\gamma-1}K + \frac{\gamma+1}{2(\gamma-1)}\log\{2(\gamma-1)\}.
\end{align}
%%2015/7/22
%We divide the Eq. (\ref{eq_differentiated_mach_number}) and (\ref{eq_mach_number}) by the unit length $r_\mathrm{0}$, we obtain the non-dimensional equations,
%\begin{align}
%& \frac{\mathcal{M}^2-1}{\mathcal{M}^2\{(\gamma-1)\mathcal{M}^2+2\}}\frac{d\mathcal{M}^2}{dx} \nonumber\\
%& \qquad = \frac{2}{x} - \frac{\gamma+1}{2(\gamma-1)} \frac{1}{1-\Phi_\mathrm{n}} \frac{d\Phi_\mathrm{n}}{dx} ,
%\end{align}
%\begin{align}
%& \mathcal{M}^{-1} \{ (\gamma-1)\mathcal{M}^2+2 \}^{\frac{\gamma+1}{2(\gamma-1)}} \nonumber\\
%& \qquad\qquad\qquad = C x^2 (1-\Phi_\mathrm{n})^{\frac{\gamma+1}{2(\gamma-1)}} ,
%\end{align}
%\begin{align}
%N(x) = \frac{4}{x} - \frac{\gamma+1}{\gamma-1} \frac{1}{1-\Phi_\mathrm{n}} \frac{d\Phi_\mathrm{n}}{dx}. \label{eq_N(x)}
%\end{align}
%where $x(=r/r_\mathrm{0})$ and $\Phi_\mathrm{n}(=\Phi/E)$ are non-dimensional radius and gravitational potential, respectively. 
%Integral constant $C$ is expressed by $K$,
%\begin{align}
%\log C = -\frac{\gamma}{\gamma-1}K + \frac{\gamma+1}{2(\gamma-1)}\log\{2(\gamma-1)\}.
%\end{align}

We adopt a model of the mass density profile of the DMH \citep{tsuchiya13} where
\begin{align}
\rho_{DMH}(r;\alpha,r_\mathrm{d},\rho_\mathrm{d}) = \frac{\rho_\mathrm{d} r_\mathrm{d}^3}{r^\alpha (r+r_\mathrm{d})^{3-\alpha}}. \label{eq_dmh_density}
\end{align}
Here, $\rho_\mathrm{d}$ represents the scale density and 0$<\alpha<$3.
In this model, we define $r_\mathrm{0}=r_\mathrm{d}$.
In the limit $x\rightarrow 0$, $\rho_\mathrm{DMH} \propto r^{-\alpha}$ and $\rho_\mathrm{DMH} \propto x^{-3}$ for $x\rightarrow \infty$.
This polytropic model reproduces various models developed from both theoretical and observational perspectives with varying degrees of accuracy.
For example, the polytropic model corresponds exactly with the NFW model \citep{navarro96} for $\alpha=1$ and approximately with the Moore model \citep{moore99,fukushige97} for $\alpha=1.5$ and the Burkert model \citep{burkert95} for $\alpha=0$.
The plausible value of the index $\alpha$ remains an open question.
Thus, we treat $\alpha$ as a variable parameter in this study.
%%2015/7/22
%We adopt a model of the mass density profile of DMH \citep{tsuchiya13} as
%\begin{align}
%\rho_{DMH}(r;\alpha,r_\mathrm{d},\rho_\mathrm{d}) = \frac{\rho_\mathrm{d} r_\mathrm{d}^3}{r^\alpha (r+r_\mathrm{d})^{3-\alpha}}. \label{eq_dmh_density}
%\end{align}
%where $\rho_\mathrm{d}$ represents the scale density and 0$<\alpha<$3. 
%In this model, we define that the unit length $r_\mathrm{0}$ is $r_\mathrm{d}$. 
%In the limit of $x\rightarrow 0$, $\rho_\mathrm{DMH} \propto r^{-\alpha}$ and $\rho_\mathrm{DMH} \propto x^{-3}$ for $x\rightarrow \infty$. 
%This model well and approximately reproduces various models predicted from both theoretical and observational perspectives. 
%For example, this model corresponds exactly to the NFW model \citep{navarro96} with $\alpha=1$ and approximately to the Moore model \citep{moore99,fukushige97} with $\alpha=1.5$ and the Burkert model \citep{burkert95} with $\alpha=0$. 
%The plausible value of the index $\alpha$ is remained as an open question. 
%Thus, we treat the concentration parameter $\alpha$ as a variable parameter in this study. 

Using Eq. (\ref{eq_dmh_density}), we obtain 
\begin{align}
& \Phi_\mathrm{n}(x;\alpha,K_\mathrm{DMH},K_\mathrm{BH}) \nonumber\\
& \quad =K_\mathrm{DMH} \int \frac{2}{x^2} \left\{ \int_0^x x^{2-\alpha} (x+1)^{\alpha-3} dx \right\} dx - K_\mathrm{BH} \frac{2}{x} ,
\end{align}
where
\begin{align}
K_\mathrm{DMH} =\frac{2\pi G \rho_\mathrm{d} r_\mathrm{d}^2}{E}, \label{eq_kdmh} \\
K_\mathrm{BH}  =\frac{GM_\mathrm{BH}}{2r_\mathrm{d}E}. \label{eq_kbh}
\end{align}
Here, $G$ and $M_\mathrm{BH}$ are the Newtonian gravitational constant and the mass of the SMBH, respectively.
The parameter $K_\mathrm{DMH}$ approximately corresponds to the ratio of the gravitational potential energy $2\pi G \rho_\mathrm{d} r_\mathrm{d}^2$ to $E$.
Similarly, the parameter $K_\mathrm{BH}$ approximately corresponds to the ratio of the gravitational potential energy $G M_\mathrm{BH} /r_\mathrm{d}$ to $E$.
Note that, in the isothermal model, $K_\mathrm{DMH}$ and $K_\mathrm{BH}$ are defined by constant $c_\mathrm{s}$ in the denominator \citep{tsuchiya13,igarashi14}. 
Here, however, we define these parameters based on $E$, as shown in Eqs. (\ref{eq_kdmh}) and (\ref{eq_kbh}).

\section{Results} \label{results}

\subsection{Transonic solutions incorporating DMH gravitational potential} \label{solutions_dmh}

\begin{figure*}
 \centering
 \includegraphics[width=2.\columnwidth]{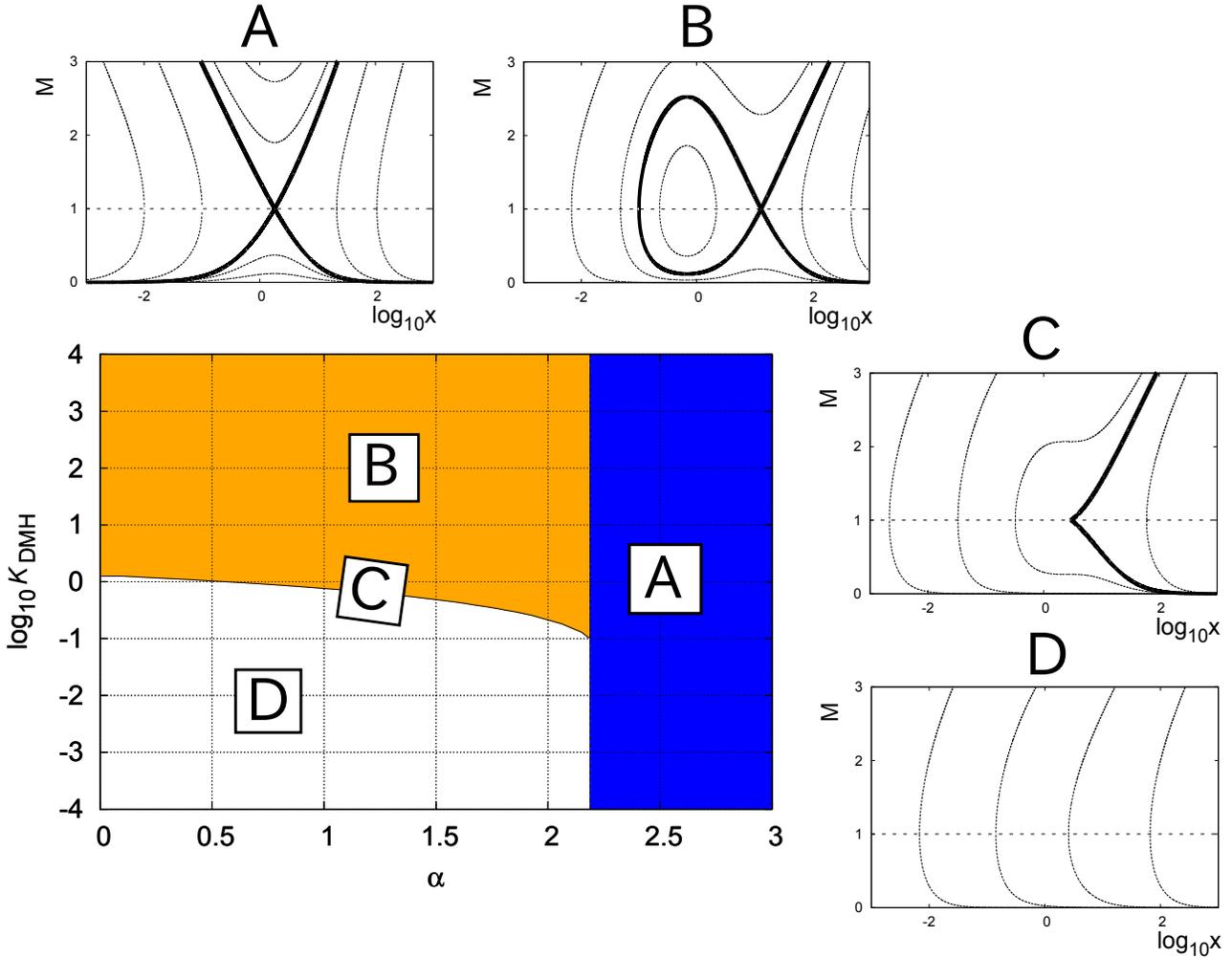}
 \caption{Various model solutions for $\gamma=1.1$. The horizontal axis shows the DMH power-law index values and the vertical axis is the $K_\mathrm{DMH}$ defined in Eq. (\ref{eq_kdmh}). Inset panels (A) and (B) represent transonic solutions, where (A) is the blue region and has only one X-point (transonic point), and (B) is the orange region has one X-point and one O-point. Panel (D) is the white region and has no critical points, while panel (C) is the boundary solution between (B) and (D). The parameters of $\mathcal{M}-x$ diagrams are chosen to clarify the topological features of transonic solutions.}
 \label{fig1}
\end{figure*}

\begin{figure*}
 \centering
 \includegraphics[width=2.\columnwidth]{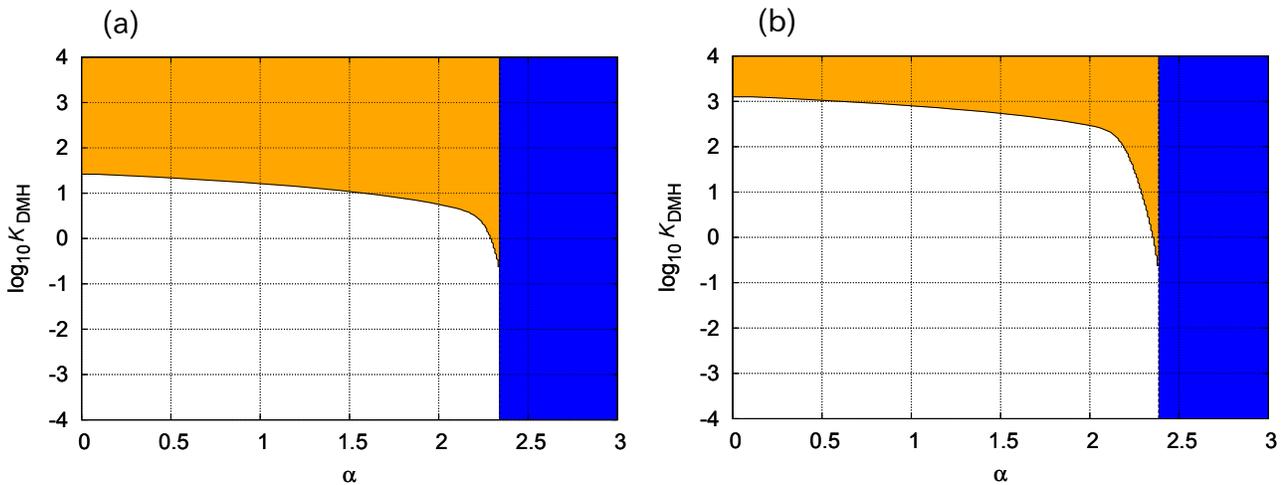}
 \caption{Solution maps for model incorporating DMH gravitational potential. Various solutions with (a) $\gamma=4/3$ and (b) $\gamma=1.5$. The colours of the divided regions correspond to those of the solution types shown in Fig. \ref{fig1}. }
 \label{fig2}
\end{figure*}

In Figs. \ref{fig1} and \ref{fig2}, we summarise the transonic solution patterns obtained when the gravitational potential of the DMH is incorporated in the model.
We identify two types of transonic solutions: One having only one X-point (the blue region labelled `A' in Fig. \ref{fig1}) and the other having one X-point with a single O-point (the orange region labelled `B' in Fig. \ref{fig1}).
The transonic solution of type A originates at the centre and extends to infinity, whereas that of type B also extends to infinity but does not originate at the centre.
These two types of transonic solutions have also been found in the isothermal model (see Figs. 2 and 4 of Tsuchiya et al. 2013 or Fig. 1 of Igarashi et al. 2014).
In the type B solutions, it is plausible to consider that the outflow starts at the vicinity of the locus of the O-point (just below the O-point of type B in Fig.\ref{fig1}).
One may feel strange that the solution does not originate at the centre. 
However, this situation is quite similar to the solar wind that starts from the corona as the material reservoir. 
In fact, the velocity is small at that point (see discussion in Igarashi et al. 2014).
Additionally, the estimated loci of the O-points for actual galaxies are close to the edge of the stellar distribution ($\sim$ several tens kpc).
This indicates that the type B solution corresponds to the slowly accelerating outflow. 
In this case, the widely-spread stellar components play a role of the reservoir of the fluid material and the energy (see Discussions \ref{Mass flux from Sombrero Galaxy} and \ref{Velocity distributions in actual galaxies}). 

When $\alpha$ is less than the critical value $\alpha_\mathrm{c}(\gamma)$ and $K_\mathrm{DMH}$ is small, there is no transonic solution (the white region labelled `D' in Fig. \ref{fig1}).
This $\alpha_\mathrm{c}(\gamma)$ value can be determined analytically using $\gamma$ (see Section \ref{alpha_value}).
When $\alpha>\alpha_\mathrm{c}(\gamma)$, type-A solutions are obtained, while type-B solutions occur when $\alpha<\alpha_\mathrm{c}(\gamma)$ with large $K_\mathrm{DMH}$.
The relationship between $\gamma$, $\alpha$, $K_\mathrm{DMH}$ and the locus of the critical point is discussed in Section \ref{critical_points}.

When $\gamma$ approaches $5/3$ (the specific heat ratio of monatomic molecule gas), the D region expands and the A and B regions contract (see Fig. \ref{fig2}).
Thus, for an adiabatic-like state ($\gamma \rightarrow 5/3$) and small $\alpha$, outflows in large galaxies correspond to type-B transonic solutions while those in small galaxies become supersonic (type D) everywhere, i.e., from the starting point to infinity.
When $\gamma=5/3$, there is no transonic solution; this is similar to the findings of previous studies \citep{mathews71,parker65}.

\subsection{Transonic solutions incorporating gravitational potentials of DMH and SMBH} \label{solutions_dmh_smbh} 

\begin{figure*}
 \centering
 \includegraphics[width=2.\columnwidth]{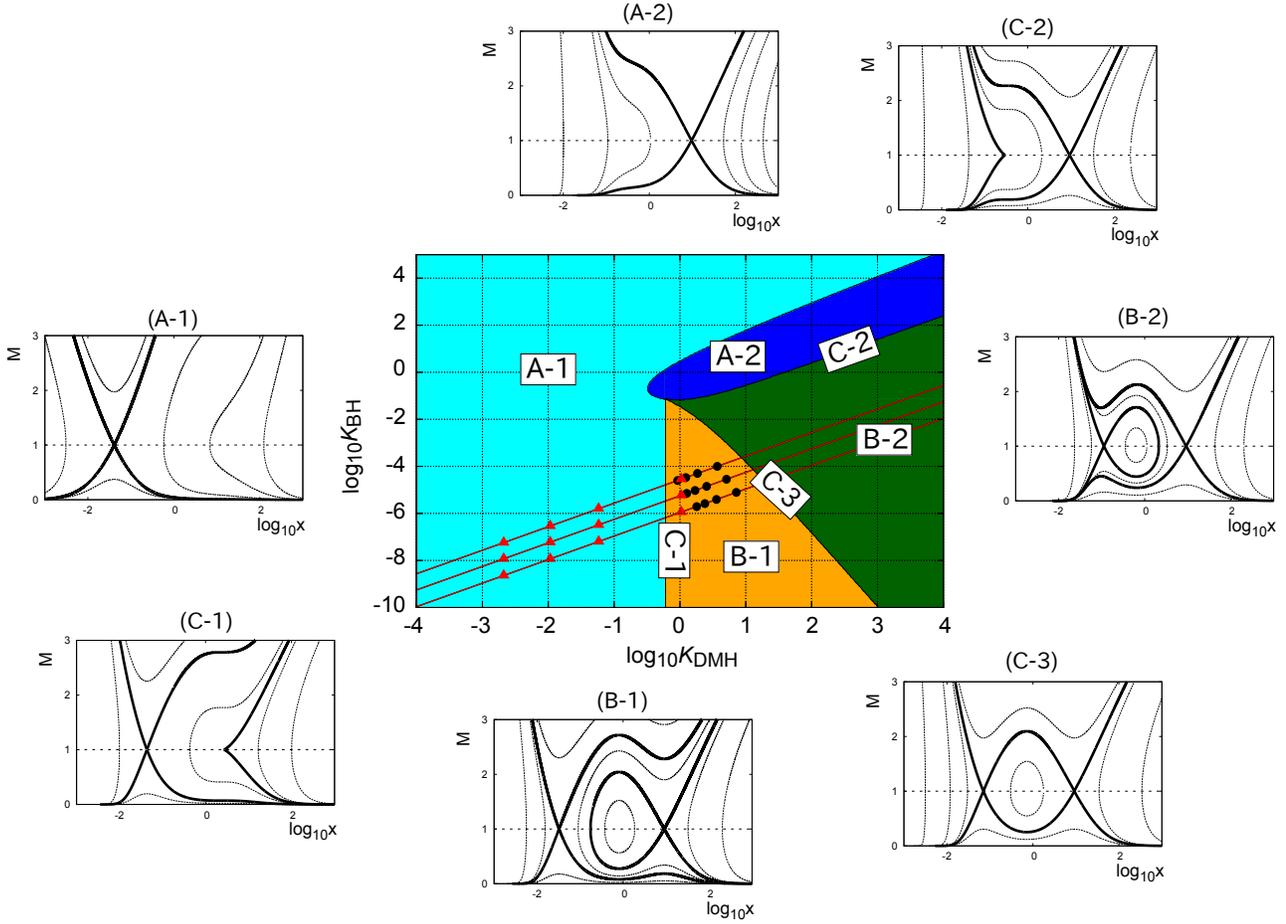}
 \caption{Various model solutions incorporating DMH and SMBH gravitational potentials for $(\gamma, \alpha)=(1.1, 1.0)$. The horizontal axis is the $K_\mathrm{DMH}$ defined in Eq. (\ref{eq_kdmh}) and roughly corresponds to the ratio of the gravitational potential energy of the DMH and $E$. 
The vertical axis is the $K_\mathrm{BH}$ defined in Eq. (\ref{eq_kbh}) and roughly corresponds to the ratio of the gravitational potential energy of the SMBH and $E$. 
The three solid lines represent the DMH mass for $10^{7}\mathrm{M_{\odot}}$, $10^{10}\mathrm{M_{\odot}}$ and $10^{13}\mathrm{M_{\odot}}$ from the bottom. 
For the quiescent galaxies, the black dots represent $\eta=1.5, 2, 2.5$, and 3, from the right. 
For the star-forming galaxies, the red dots represent $\eta'=1.5, 10, 50$, and 250, from the right. 
See Section \ref{parameters} for details. The parameters of $\mathcal{M}-x$ diagrams are chosen to clarify the topological features of transonic solutions.}
 \label{fig3}
\end{figure*}

\begin{figure*}
 \centering
 \includegraphics[width=2.\columnwidth]{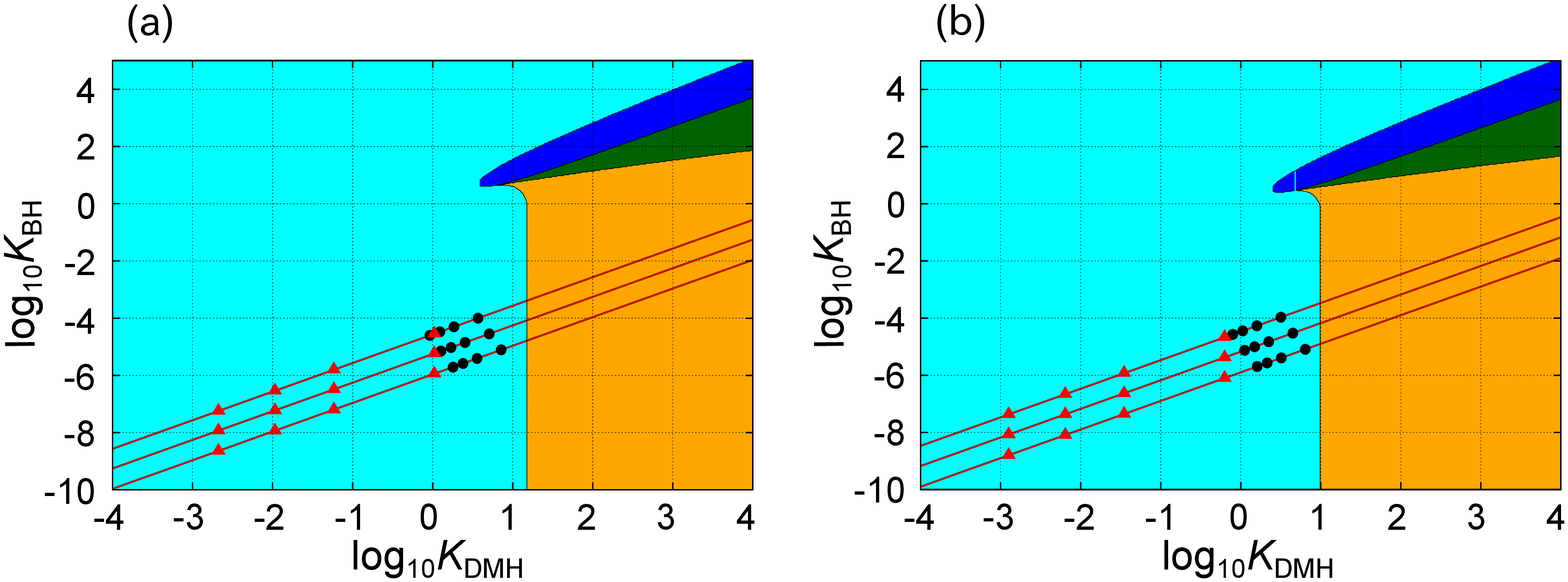}
 \caption{Solution maps for model incorporating DMH and SMBH gravitational potentials. Various solutions with (a) $(\gamma,\alpha)=(4/3,1.0)$ and (b) $(\gamma,\alpha)=(4/3,1.5)$. The colours of the divided regions correspond to the solutions types given in Fig. \ref{fig3}. The red and black lines also correspond to the galaxy parameters given in Fig. \ref{fig3}. }
 \label{fig4}
\end{figure*}

In this section, we add the gravitational potential of the SMBH to that of the DMH in the model.
The DMH potential is widely distributed and dominant over a large region within a given galaxy.
In contrast, the potential of the SMBH is dominant in the vicinity of the centre only.
Hence, the gravitational potential of the SMBH is important as regards transonic solutions originating at the centre.
We summarise the subtypes of the transonic solutions obtained while incorporating the gravitational potentials of both the DMH and SMBH in the model in Figs. \ref{fig3} and \ref{fig4}.
As in \citet{igarashi14}, the gravitational potential of the SMBH adds a new branch to the transonic solutions yielded by the model discussed previously, which incorporated the effects of the DMH gravitational potential only.
We determine two primary types of transonic solution: One having only one X-point (type A) and the other having two X-points with a single O-point (type B).

In the B case, the inner/outer X-points are due to the SMBH/DMH potentials, respectively.
The transonic solution through the inner/outer X-point is referred to as type $\rmn{X_{in}/X_{out}}$, as in \citet{igarashi14}.
We can topologically divide the B case into two sub-types: B-1 and B-2.
In the B-1 region (the orange region labelled `B-1' in Fig. \ref{fig3}), the type-$\rmn{X_{in}}$ solution originates at the centre, but the type-$\rmn{X_{out}}$ solution does not.
Both solutions extend to infinity.
In the B-2 region (the green region labelled `B-2' in Fig. \ref{fig3}), the type-$\rmn{X_{out}}$ solution extends to infinity, but the type-$\rmn{X_{in}}$ solution does not.
Both solutions originate at the centre.
On the boundary between regions B-1 and B-2, a special transonic solution connecting two X-points appears (the line labelled `C-3' in Fig. \ref{fig3}).

In the isothermal model \citep{igarashi14}, the type-A solutions are divided into three subtypes depending on the loci of the extreme points of $N(x)$. 
In this paper, we divide the type-A solutions into two subtypes, A-1 and A-2, based on a comparison of the gravity $d\Phi/dx$ of the DMH and SMBH.
In the A-1 case (cyan region labelled `A-1' in Fig. \ref{fig3}), the gravity of the SMBH is greater than that of the DMH at the transonic point, and in the A-2 case (blue region labelled `A-2' in Fig. \ref{fig3}), that of the DMH is greater than that of the SMBH.
%2015/7/22
%In the isothermal model \citep{igarashi14}, the type A solutions were divided into three patterns depending on the locus of extreme points of $N(x)$. In this paper, we divide type A solutions into two patterns, A-1 and A-2, using the comparison of gravity $d\phi/dx$ of DMH and SMBH. 
%In type A-1 (cyan region labeled as type A-1 in Fig. \ref{fig3}), the gravity of SMBH is greater than that of DMH at the transonic point, and in type A-2 (blue region labeled as type A-2 in Fig. \ref{fig3}), that of DMH is greater than that of SMBH. 
%%2015/4/14
%In the isothermal model \citep{igarashi14}, type A solutions were devided into three patterns depending on the locus of extreme points of $N(x)$, while We divide type A solutions into two patterns, A-1 and A-2, using the comparison of gravity $d\phi/dx$ of DMH and SMBH in this paper. 
%In type A-1 (cyan region labeled as type A-1 in Fig. \ref{fig3}), the gravity of SMBH is more dominant than that of DMH at the transonic point, while in type A-2 (blue region labeled as type A-2 in Fig. \ref{fig3}), that of DMH is more dominant than that of SMBH. 

When $K_\mathrm{DMH}$ and $K_\mathrm{BH}$ are large and small, respectively, the outflow solutions are of type B-1 and B-2.
Specifically, with very large $K_\mathrm{DMH}$ (orange region in Figs. \ref{fig3} and \ref{fig4}), the outflow solutions become type B-2.
Further, because the type-$\rmn{X_{in}}$ solution in the B-2 case does not extend to infinity, only the type-$\rmn{X_{out}}$ solution is available for the transonic outflow.
When both $K_\mathrm{BH}$ and $K_\mathrm{DMH}$ are large, the solution becomes type A-2.
Furthermore, when $K_\mathrm{BH}$ is very large (cyan region in Figs. \ref{fig3} and \ref{fig4}), the solution becomes type A-1.
With small $K_\mathrm{DMH}$, the solution is type A-1, because the gravity of the SMBH is greater than that of the DMH in the vicinity of centre.
%2015/7/22
%When $K_\mathrm{DMH}$ is large with small $K_\mathrm{BH}$, the solutions of outflows become type B-1 and B-2. 
%Specifically, with very large $K_\mathrm{DMH}$, the solution of outflows becomes type B-2. 
%Because type $\rmn{X_{in}}$ solution in type B-2 does not extend to infinity, only the type $\rmn{X_{out}}$ solution is available for the transonic outflow in type B-2. 
%When $K_\mathrm{BH}$ is large with large $K_\mathrm{DMH}$, the solution becomes type A-2. 
%Furthermore, when $K_\mathrm{BH}$ is very large, the solution becomes type A-1. 
%With small $K_\mathrm{DMH}$, the solution is type A-1, because the gravity of SMBH is greater than that of DMH at the vicinity of centre. 
%2015/4/14
%When $K_\mathrm{DMH}$ is large with small $K_\mathrm{BH}$, the solution of outflows becomes type B-1 and B-2. 
%Specifically, with very large $K_\mathrm{DMH}$, the solution of outflows becomes type B-2. 
%Because type $\rmn{X_{in}}$ solution in type B-2 does not extends to infinity, the transonic outflow in type B-2 is type $\rmn{X_{out}}$ solution only. 
%When $K_\mathrm{BH}$ is large with large $K_\mathrm{DMH}$, the solution becomes type A-2. 
%Furthermore, when $K_\mathrm{BH}$ is more large, the solution becomes type A-1. 
%With small $K_\mathrm{DMH}$, the solution is type A-1 because the gravity of SMBH is more dominant than that of DMH at the vicinity of centre. 

When $\gamma$ is close to $5/3$, the A-1 region extends while the A-2, B-1, and B-2 regions contract (see Figs. \ref{fig3} and \ref{fig4}).
This result indicates that the transonic point of the transonic outflow in the adiabatic-like state is located in the inner region.
When $\alpha$ is large with small $K_\mathrm{BH}$, the B-1 and B-2 regions are extended and the A-1 and A-2 regions contract.
In contrast, when $\alpha$ is large with large $K_\mathrm{BH}$, the A-1 and A-2 regions extend and the B-1 and B-2 regions contract.
When $\alpha$ is larger than the critical value $\alpha_\mathrm{c}(\gamma)$, only solutions of type A-1 and A-2 occur, i.e., solutions of type B-1 and B-2 do not appear.
This $\alpha_\mathrm{c}(\gamma)$ has the same value as in the case where the DMH potential only is considered (see Section \ref{alpha_value}).
Further, the dependence of $\alpha_\mathrm{c}(\gamma)$ on $\gamma$ is examined in Section \ref{alpha_value}.
Note that, when $\gamma=5/3$, there is no transonic solution; this is also similar to the case in which the DMH potential only is considered.

%%2015/7/22
%\begin{figure*}
% \centering
% \includegraphics[width=2.\columnwidth]{fig1.eps}
% \caption{Various solutions of the model with $\gamma=3.3/3.0$. The horizontal axis shows the power-law index of DMH, the vertical axis is $K_\mathrm{DMH}$ defined in Eq. (\ref{eq_kdmh}). The inset panels (A) and (B) represent transonic solutions. The panel (A) in blue region has only one X-point (transonic point), (B) in the orange region has one X-point and one O-point. The panel (D) in white region has no critical points. The panel (C) is the boundary solution between (B) and (D). }
% \label{fig1}
%\end{figure*}

%%2015/7/22
%\begin{figure*}
% \centering
% \includegraphics[width=2.\columnwidth]{fig2.eps}
% \caption{Solution maps of the model with dark matter halo. Various solutions with $\gamma=4.0/3.0$ are represented in (a) that with $\gamma=4.5/3.0$ shown in (b). The color of the divided regions corresponds to types of solutions as in Fig. \ref{fig1}. }
% \label{fig2}
%\end{figure*}

%%2015/7/22
%\begin{figure*}
% \centering
% \includegraphics[width=2.\columnwidth]{fig3.eps}
% \caption{Various solutions of the model with dark matter halo (DMH) and the central black hole (SMBH) with $(\gamma,\alpha)=(3.3/3.0,1.0)$. The horizontal axis is $K_\mathrm{DMH}$ defined in Eq. (\ref{eq_kdmh}) and roughly corresponds to the ratio of the gravitational potential energy of DMH and total energy. The vertical axis is $K_\mathrm{BH}$ defined in Eq. (\ref{eq_kbh}) and roughly corresponds to the ratio of the  gravitational potential energy of SMBH and total energy. The three black lines represent parameters of slowly accelerating outflows from $10^{11}\mathrm{M_{\odot}}$ (bottom) to $10^{13}\mathrm{M_{\odot}}$ (top). The black dots show the correction parameter $\eta$ is 1.5, 2, 2.5 and 3 from right. The three red lines represent parameters of high-velocity outflows from $10^{11}\mathrm{M_{\odot}}$ (bottom) to $10^{13}\mathrm{M_{\odot}}$(top). The red dots show the correction parameter $\eta'$ is 2, 10, 50 and 250 from right. See Section \ref{parameters} for details. }
% \label{fig3}
%\end{figure*}

%%2015/7/22
%\begin{figure*}
% \centering
% \includegraphics[width=2.\columnwidth]{fig4.eps}
% \caption{Solution maps of the model with dark matter halo and the central black hole. Various solutions with $(\gamma,\alpha)=(4.0/3.0,1.0)$ are represented in (a) that with $(\gamma,\alpha)=(4.5/3.0,1.5)$ shown in (b). The color of he divided regions corresponds to types of solutions as in Fig. \ref{fig3}. The red and black lines also correspond to parameters of galaxies as in Fig. \ref{fig3}. }
% \label{fig4}
%\end{figure*}

in cases A, B and D, the function f(x) diverges as $x \rightarrow 0$, i.e. at the centre of the dark matter distribution
\subsection{Critical value of $\alpha$} \label{alpha_value}

In Section \ref{solutions_dmh}, we found that a single X-point with no O-point (type-A region in Fig. \ref{fig1}) occurs when $\alpha$ is larger than $\alpha_\mathrm{c}(\gamma)$.
This critical value depends on $\gamma$.
In this section, we estimate $\alpha_\mathrm{c}(\gamma)$ using $f(x)$ $(=N(x)/2)$, which is defined by Eq. (\ref{eq_N(x)}).
In cases A, B and D, the function $f(x)$ diverges as $x \rightarrow 0$, i.e. at the centre of the dark matter distribution.

First, we integrate Eq. (\ref{eq_N(x)}) and obtain
\begin{align}
F(x) &= \int f(x) dx, \nonumber\\
     &= 2\log x + \frac{\gamma+1}{2(\gamma-1)}\log(1-\Phi_\mathrm{n}).
\end{align}
When $2< \alpha <3$, we can expand the exponent of $F(x)$ at the centre, such that
\begin{align}
& \exp\{F(x)\} \nonumber\\
& \qquad = x^2 (1-\Phi_\mathrm{n})^\frac{\gamma+1}{2(\gamma-1)}, \nonumber\\
& \qquad \approx x^2 \left\{ 1-2K_\mathrm{DMH}\frac{x^{2-\alpha}}{(\alpha-2)(\alpha-3)} +O[x^{3-\alpha}] \right\}^\frac{\gamma+1}{2(\gamma-1)}.
\end{align}
Taking the limit $x\rightarrow 0$ of the above expression, we find
\begin{align}
\exp\{F(x)\} \propto x^{2+\frac{\gamma+1}{2(\gamma-1)}(2-\alpha)}.
\end{align}
When the exponent of $x$ becomes 0, the transonic solution types change from B and D to A.
Thus, we obtain 
\begin{align}
\alpha_\mathrm{c} &= \frac{2(3\gamma-1)}{\gamma+1}, \label{eq_critical_alpha}
\end{align}
where $\alpha_\mathrm{c}$ depends on $\gamma$.
This equation clearly indicates that $\gamma$ and $\alpha$ are essential to determine the type of galactic outflows.
The solution through the outer transonic point needs the specific balance of the supplied thermal energy from stellar components and the structure of gravitational potential.
In the isothermal model, \citet{tsuchiya13} indicated that $\alpha_\mathrm{c}=2$, and
this result can be reproduced here by substituting $\gamma=1$ into Eq. (\ref{eq_critical_alpha}).
The results in Fig. \ref{fig1} are consistent with this analysis.

When the gravitational potential of the SMBH is incorporated into the model, $\alpha_\mathrm{c}(\gamma)$ can also be obtained through this analysis.
When $\alpha<\alpha_\mathrm{c}$, solutions of type A-1, A-2, B-1, and B-2 occur (see Fig. \ref{fig2}).
However, when $\alpha>\alpha_\mathrm{c}$, the solutions of type B-1 and B-2 are absent.
%2015/7/22
%When the gravitational potential with SMBH, the critical $\alpha_\mathrm{c}(\gamma)$ is also represented as this analysis. 
%When $\alpha<\alpha_\mathrm{c}$, there are type A-1 and A-2 solutions with type B-1 and B-2 solutions (see Fig. \ref{fig2}). 
%When $\alpha>\alpha_\mathrm{c}$, there are type A-1 and A-2 solutions without type B-1 and B-2 solutions. 
%%2015/4/14
%When the gravitational potential contains that of SMBH, the critical $\alpha$ is also represented as this analysis. 
%When $\alpha$ is larger than the value indicated by Eq. (\ref{eq_critical_alpha}), there is single X-point (type A-1 and A-2 in Fig. \ref{fig2}) not including two X-pints with single O-point (type B-1 and B-2). 

%2014/11/11
%\begin{itemize}
%\item classification with positive energy in DMH (Figs. \ref{fig1}, \ref{fig2} \& \ref{fig3})
%\item re-definition of solution pattern in DMH and SMBH
%\item classification with positive energy in DMH and SMBH (Figs. \ref{fig4}, \ref{fig5}, \ref{fig6} \& \ref{fig7})
%\item influence of $\gamma$
%\item relation of $\alpha$ and $\gamma$ in critical solutions
%\end{itemize}

\section{Discussion} \label{discussion}

\subsection{Inquiry into Assumptions}\label{validity of assumptions}
In this study, we assume steady, spherically symmetric and polytropic galactic outflows ignoring mass and energy injections along flow lines.
We discuss the rationality of these assumptions in this section. 
%%2016/06/27
%\textcolor{red}{
%\subsection{Validity of Assumptions}\label{validity of assumptions}
%In this study, we assume steady, spherically symmetric and polytropic galactic outflows ignoring mass and energy injections along flow lines.
%We discuss the validity of these assumptions in this section. 
%}
%%2015/10/16
%\textcolor{red}{
%\subsection{Availability of Assumptions}\label{availability of assumptions}
%In this study, we assume steady, spherically symmetric and polytropic galactic outflows ignoring mass and energy injections along flow lines.
%We investigate the availability of these assumptions in this section.}

The steady assumption is available for the case in which the wind crossing timescale is much shorter than the timescale for the temporal variation of the energy supply from the stellar system. 
In addition, we must note that the steady assumption is also available for the case in which the wind crossing timescale is much longer than the averaged interval of the frequent energy supply. 
For slowly accelerating galactic outflows, the crossing timescale is $r_\mathrm{d}/v$ ($\sim 10 \mathrm{kpc}/100\mathrm{km}\,\mathrm{s^{-1}}\sim 100\mathrm{Myr}$).
The prime driving force of the slowly-accelerating outflows is probably thought to be Type Ia supernovae and stellar winds.
In this case, the energy supply is regulated by the stellar evolution and is a secular process with an almost constant frequency and a feedback energy.
Under these situations, the steady state is a reasonable approximation.
On the other hand, for the galactic winds from star-forming galaxies, the source of the energy supply is mainly massive stars.
In this case, the typical timescale of the energy supply and the crossing timescale have wide variations, and the steady assumption may often be violated.

Because the spatial distribution of galactic outflows observed in star-forming galaxies is known to be multidimensional \citep{martin06,mori06}, an analytical approach is tough for such complicated structure.
Recent theoretical analyses take advantage of numerical studies for this reason.
However, the systemic and fundamental comprehension on the transonic galactic outflows in the actual gravitational potential model has not been accomplished yet. 
Thus, we assume here steady spherically symmetric outflows as the simplest model to clarify the fundamental nature of the transonic galactic outflows in this paper.
Since the non-spherical component such as a stellar disc will affect the topology of the transonic solutions, it is interesting to classify and summarize the variety of transonic solutions with the non-spherical components.

The mass and the energy injections from stellar winds and supernovae will act as braking or accelerating process in actual galactic outflows.
We note that since these injections are relevant only in the stellar distribution region, we can ignore them for the widely spread acceleration region if the transonic point forms in outside the stellar distribution region.
On the other hand, the mass and the energy injections may influence the loci of the transonic points of the high-velocity galactic outflows, because the transonic points for such flows are located in the stellar distribution region. 
Thus, we leave the details of this discussion including the stability of transonic solutions to our future study. 

In addition, we ignore the stellar component as a gravitational source.
In our previous paper (see Sec. 4.4.2 of Igarashi et al. 2014), we applied the isothermal model to the Sombrero galaxy and found that the stellar gravity does not strongly influence the acceleration process of the slowly accelerating galactic outflow which will be relevant for this galaxy. 
This result clearly shows that the effect of the DMH gravity is dominant comparing to the effect of the stellar gravity for normal galaxies ($\gtrsim 10^\mathrm{10}M_\mathrm{\odot}$).
Thus, it seems reasonable to ignore the effect of the stellar gravity for the normal galaxies in this study.
In our future work, we plan to investigate the transonic outflows in the gravitational potential with the stellar mass component for the normal galaxies. 

\subsection{Relation between parameters and critical point loci} \label{critical_points}

\begin{figure*}
 \centering
 \includegraphics[width=1.3333\columnwidth]{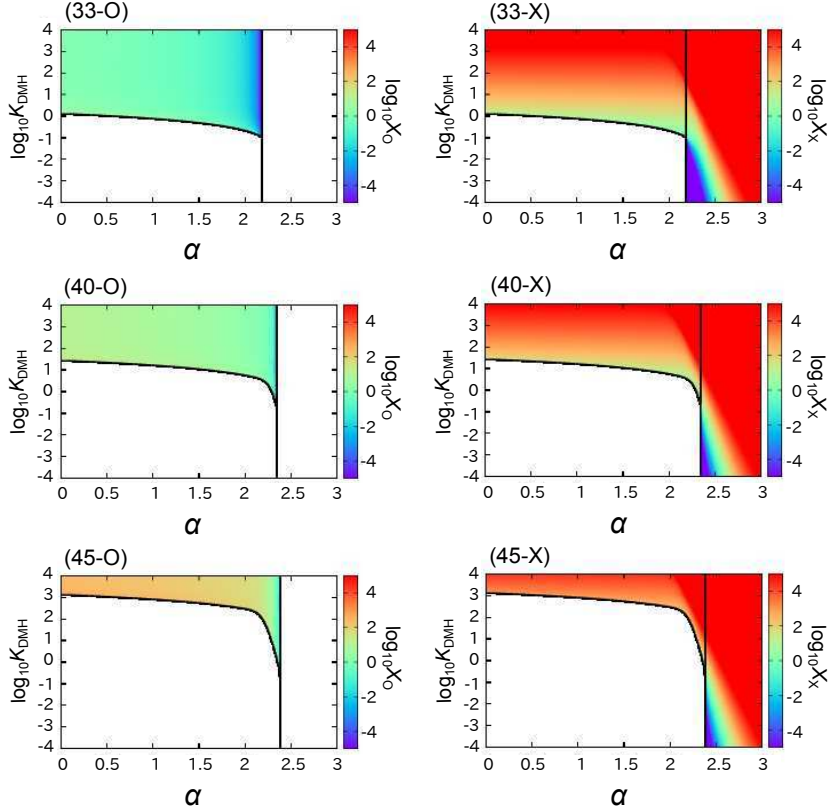}
 \caption{Loci of critical points for the model incorporating DMH gravitational potential. The labels above the graphs represent the values of the $\gamma$ parameter and the critical point types. For example, the graphs labelled (33-O) and (45-X) show the loci of an O-point with $\gamma=1.1$ and an X-point with $\gamma=1.5$, respectively. $X_\mathrm{X}$ and $X_\mathrm{O}$ denote the loci of the X-point and the O-point, respectively. Color bars correspond to the loci of these critical points in the logarithmic scale.}
 \label{fig5}
\end{figure*}

In Fig. \ref{fig5}, we show the loci of critical points obtained when the gravitational potential of the DMH is incorporated into the model.
$X_\mathrm{X}$ and $X_\mathrm{O}$ in Fig. \ref{fig5} denote the loci of the X-point and the O-point, respectively. 
Each colour bar shows the loci of these critical points in the logarithmic scale.
When either the $\alpha$ of the DMH or the $K_\mathrm{DMH}$ increases, the positions of the X- and O-points move outward and inward, respectively.
In contrast, when $\gamma$ increases, the position of the X-point moves inward and that of the O-point moves outward.
This change in the behaviour of the critical points indicates that the value of $\gamma$ is critical to the acceleration process of galactic outflows.
Because the observed thermal distribution varies with heating by stars and radiative cooling \citep{fukuzawa06,diehl08} and $\gamma$ is dependent on these processes, $\gamma$ is variable in actual galaxies.
Therefore, heating and cooling in the interstellar gas influences the acceleration process of the actual galactic outflows.

\begin{figure*}
 \centering
 \includegraphics[width=2.\columnwidth]{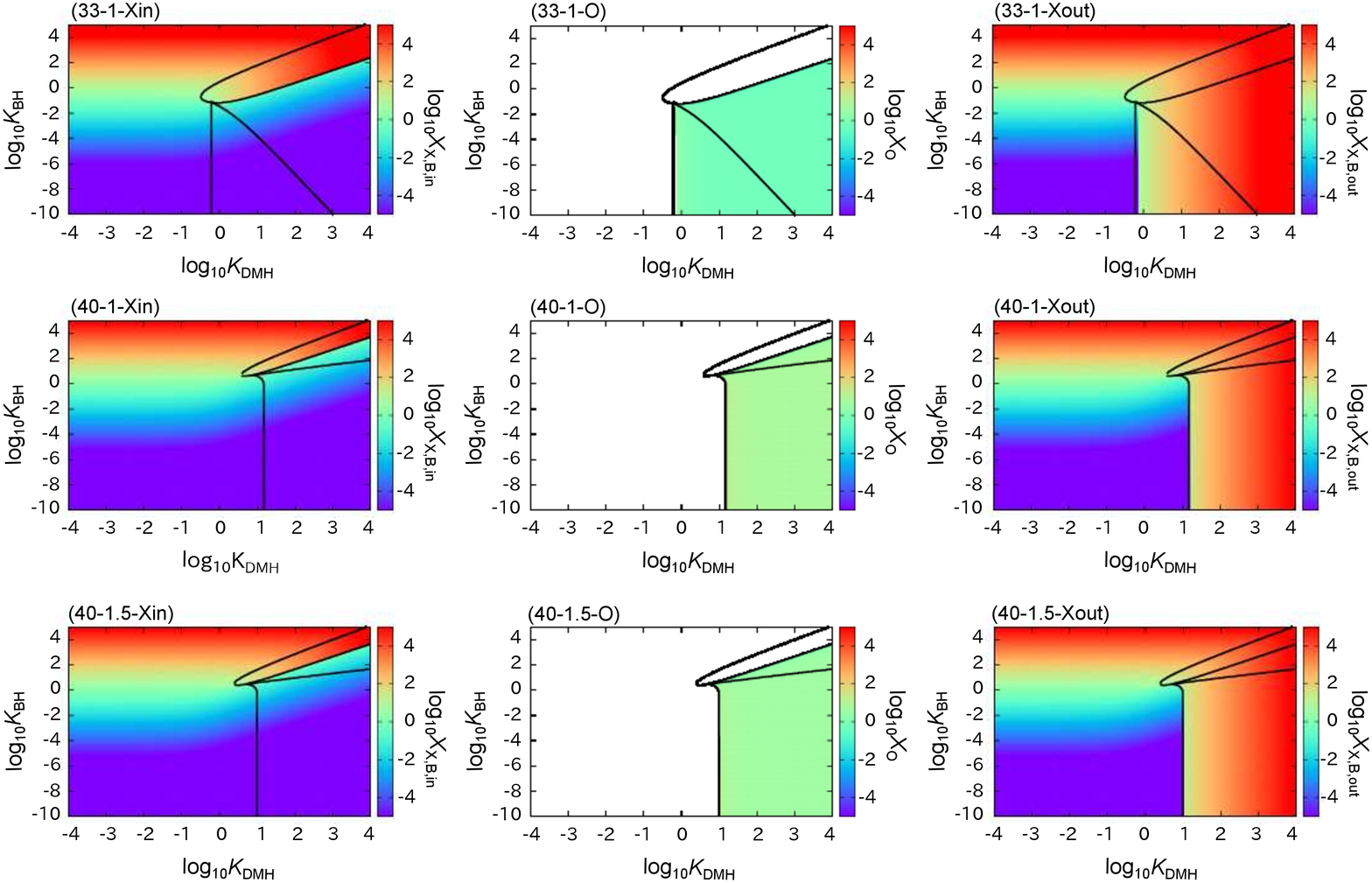}
 \caption{Loci of critical points for the model incorporating DMH and SMBH gravitational potentials. The labels above the graphs represent the values of the parameters ($\gamma,\alpha$) and the critical point types. For example, the graphs labelled (33-1-Xin) and (40-1.5-Xout) show the loci of the inner X-point with $(\gamma,\alpha)=(1.1,1.0)$ and the outer X-point with $(\gamma,\alpha)=(4/3,1.5)$. The loci of the X-points in the A-1 and A-2 regions are shown in the $X_\mathrm{in}$ and $X_\mathrm{out}$ graphs. $X_\mathrm{X,B,in}$, $X_\mathrm{X,B,out}$ and $X_\mathrm{O}$ denote the loci of the inner X-point, the outer X-point and the O-point, respectively. Color bars correspond to the loci of these critical points in the logarithmic scale.}
 \label{fig6}
\end{figure*}

In Fig. \ref{fig6}, we show the loci of the critical points when the gravitational potentials of both the DMH and SMBH are incorporated in the model.
$X_\mathrm{X,B,in}$, $X_\mathrm{X,B,out}$ and $X_\mathrm{O}$ in Fig. \ref{fig6} denote the loci of the inner X-point, the outer X-point and the O-point, respectively. 
%%2015/10/16
%In Fig. \ref{fig6}, we show the loci of the critical points when the gravitational potentials of both the DMH and SMBH are incorporated in the model.
%\textcolor{red}{
%$X_\mathrm{X,B,in}$, $X_\mathrm{X,B,out}$ and $X_\mathrm{O}$ in Fig. \ref{fig6} represent the locus of an inner X-point, that of an outer X-point and that of an O-point, respectively. 
%The color bars next to the graphs are the loci of these critical points in the logarithmic scale.}
In the A-1 region, the position of the X-point moves outward when $\gamma$ decreases or $K_\mathrm{BH}$ increases.
This position also weakly depends on $K_\mathrm{DMH}$.
In the A-2 case, the X-point moves outward when $\gamma$ decreases or $K_\mathrm{DMH}$ increases.
This position weakly depends on $K_\mathrm{BH}$.
In the cases B-1 and B-2, the position of the inner X-point depends on $\gamma$, $K_\mathrm{DMH}$, and $K_\mathrm{BH}$, while the position of the outer X-point depends on $\gamma$ and $K_\mathrm{DMH}$ (not on $K_\mathrm{BH}$, see Fig. \ref{fig6}).
The position of the O-point of these solutions primarily depends on $\gamma$, $K_\mathrm{DMH}$ and $\alpha$, but the dependence on $K_\mathrm{DMH}$ and $\alpha$ is very weak.
Because the positions of the O-point and the outer X-point are almost independent of $K_\mathrm{BH}$, those positions in Fig. \ref{fig6} (the case with the SMBH gravity) and Fig. \ref{fig5} (the case without he SMBH gravity) become the same.
As in the case where the gravitational potential of the DMH only is considered, these results indicate that $\gamma$ influences the acceleration process of the outflows significantly, when the gravitational effects of both the DMH and SMBH are incorporated in the model.

%%2015/10/16
%\begin{figure*}
% \centering
% \includegraphics[width=1.3333\columnwidth]{fig5.eps}
% \caption{Loci of critical points for the model incorporating DMH gravitational potential. The labels above the graphs represent the values of the $\gamma$ parameter and the critical point types. For example, the graphs labelled (33-O) and (45-X) show the loci of an O-point with $\gamma=1.1$ and an X-point with $\gamma=1.5$, respectively. \textcolor{red}{$X_\mathrm{X}$ and $X_\mathrm{O}$ represent the locus of a X-point and that of an O-point, respectively. Color bars are the loci of these critical points in the logarithmic scale.}}
% \label{fig5}
%\end{figure*}
%%2015/8/3
%\begin{figure*}
% \centering
% \includegraphics[width=1.3333\columnwidth]{fig5.eps}
% \caption{Critical point loci for model incorporating DMH gravitational potential. The labels above the graphs represent the values of the $\gamma$ parameter and the critical point types. For example, the graphs labelled (33-O) and (45-X) show the loci of an O-point with $\gamma=3.3/3.0$ and an X-point with $\gamma=4.5/3.0$, respectively. }
% \label{fig5}
%\end{figure*}
%%2015/7/22
%\begin{figure*}
% \centering
% \includegraphics[width=1.3333\columnwidth]{fig5.eps}
% \caption{The locus of critical points with the model in the gravitational potential of dark matter halo. The label above the graph represents parameter $\gamma$ and type of critical points. For example, the graph (33-O) shows the locus of O-point with $\gamma=3.3/3.0$ and (45-X) shows the locus of X-point with $\gamma=4.5/3.0$. }
% \label{fig5}
%\end{figure*}

\subsection{Parameter ranges of actual galaxies}\label{parameters}

In this section, we estimate the parameter ranges of actual galaxies.
Similar to \citet{igarashi14}, we focus on the parameter ranges of slowly accelerating cases in quiescent galaxies.
Additionally, because many of the previous studies of galactic outflows primarily focused on high-velocity outflows driven by starbursts, we also estimate the parameter ranges of high-velocity outflows in active star-forming galaxies.

For the slowly accelerating outflows in quiescent galaxies, we adopt the same assumption as \citet{igarashi14} in order to estimate the parameter ranges: The physical state of the subsonic region is similar to the equilibrium state, and the subsonic region spreads up to near the virial radius $r_\mathrm{vir}$ in slowly accelerating outflows.
Therefore, we can expect that the thermal and kinetic energies of the slowly accelerating outflows near $r_\mathrm{vir}$ are slightly larger than the gravitational potential energy, and we adopt a new correction parameter, $\eta$. 
Under this assumption, Eq. (\ref{eq_energy}) becomes

\begin{align}
E &= \left.\left(\frac{c_\mathrm{s}^2}{\gamma-1}+\frac{v^2}{2}\right)\right|_\mathrm{r=r_\mathrm{vir}} + \Phi(r_\mathrm{vir}), \\
&=\eta\left|\Phi(r_\mathrm{vir})\right| + \Phi(r_\mathrm{vir}), \label{eq_slow_22} \\
&\approx (\eta-1)\left|\left(\left.\int\frac{GM_\mathrm{DMH}(r_\mathrm{vir})}{r^2}dr\right|_\mathrm{r=r_\mathrm{vir}}\right)\right|. \label{eq_slow}
\end{align}
In addition, $\eta$ is slightly larger than unity, because the sum of the thermal and kinetic energies is slightly larger than the gravitational potential energy in the slowly accelerating outflows.
In Eq. (\ref{eq_slow}), we ignore the influence of the SMBH because commonly it is smaller than that of the DMH at $r_\mathrm{vir}$.
From cosmological simulations, the relation between the DMH mass and the concentration parameter $r_\mathrm{vir}/r_\mathrm{d}$ is modelled as 

\begin{align}
\frac{r_\mathrm{vir}}{r_\mathrm{d}}=\kappa\left( \frac{M_\mathrm{DMH}(r_\mathrm{vir})}{10^{12}\mathrm{M_{\odot}}} \right)^\mathrm{\xi}. \label{concentration_parameter}
\end{align}
We adopt the parameters proposed by \citet{prada12}, $(\kappa, \xi)=(9.7, -0.074)$ (see also, \citet{navarro96,maccio08,klypin11,ogiya14}).
For $r_\mathrm{vir}$, we use the formula given by \citet{bullock01}.
We also adopt the observational relation between the masses of the SMBH and DMH, where 
%%2015/10/28
%\begin{align}
%\frac{r_\mathrm{vir}}{r_\mathrm{d}}=\kappa\left( \frac{M_\mathrm{DMH}(r_\mathrm{vir})}{10^{12}\mathrm{M_{\odot}}} \right)^\mathrm{\xi}. \label{concentration_parameter}
%\end{align}
%We adopt the parameters proposed by \citet{prada12}, $(\kappa, \xi)=(9.7, -0.074)$ (see also, \citet{navarro96,maccio08,klypin11,ogiya14}).
%\textcolor{red}{
%For $r_\mathrm{vir}$, we use results by \citet{bullock01}.}
%We also adopt the observational relation between the masses of the SMBH and DMH, where 
%%2015/8/3
%\begin{align}
%c=\kappa\left( \frac{M_\mathrm{DMH}(r_\mathrm{vir})}{10^{12}\mathrm{M_{\odot}}} \right)^\mathrm{\xi}. \label{concentration_parameter}
%\end{align}
%We adopt the parameters proposed by \citet{prada12}, $(\kappa, \xi)=(9.7, -0.074)$ (see also, \citet{navarro96,bullock01,maccio08,klypin11}).
%We also adopt the observational relation between the masses of the SMBH and DMH, where 
%%2015/7/22
%\begin{align}
%c=\kappa\left( \frac{M_\mathrm{DMH}(r_\mathrm{vir})}{10^{12}\mathrm{M_{\odot}}} \right)^\mathrm{\xi}. \label{concentration_parameter}
%\end{align}
%We adopt parameters proposed by \citet{prada12}, $(\kappa,\xi)=(9.7,-0.074)$ (see also \citet{navarro96,bullock01,maccio08,klypin11}). 
%We also adopt the observational relation of SMBH mass and DMH mass 
%%2015/3/17
%We adopted parameters proposed by \citet{prada12}, $(\kappa,\xi)=(9.7,-0.074)$ (see also \citet{navarro96,bullock01,maccio08,klypin11}). 
%Moreover, we use the observational relation of SMBH mass and DMH mass,

\begin{align}
\frac{M_\mathrm{BH}}{10^{8}\mathrm{M_{\odot}}} = \mu\left(\frac{M_\mathrm{DMH}(r_\mathrm{vir})}{10^{12}\mathrm{M_{\odot}}}\right)^\mathrm{\nu}. \label{SMBH_mass}
\end{align}
In addition, we adopt the parameters proposed by \citet{baes03}, $(\mu,\nu)=(0.11,1.27)$ (see also, Ferrarese 2002).
Finally, three driving parameters ($\gamma$, $M_\mathrm{DMH}(r_\mathrm{vir})$, $\eta$) for the slowly accelerating outflows remain.
%%2015/8/3
%\begin{align}
%\frac{M_\mathrm{BH}}{10^{8}\mathrm{M_{\odot}}} = \mu\left(\frac{M_\mathrm{DMH}(r_\mathrm{vir})}{10^{12}\mathrm{M_{\odot}}}\right)^\mathrm{\nu}. \label{SMBH_mass}
%\end{align}
%In addition, we adopt the parameters proposed by \citet{baes03}, $(\mu,\nu)=(0.11,1.27)$ (see also, \citet{ferrarese02}).
%Finally, three undetermined parameters ($\gamma$, $M_\mathrm{DMH}(r_\mathrm{vir})$, $\eta$) for the slowly accelerating outflows remain.
%%2015/7/22
%\begin{align}
%\frac{M_\mathrm{BH}}{10^{8}\mathrm{M_{\odot}}} = \mu\left(\frac{M_\mathrm{DMH}(r_\mathrm{vir})}{10^{12}\mathrm{M_{\odot}}}\right)^\mathrm{\nu}. \label{SMBH_mass}
%\end{align}
%We adopt parameters proposed by \citet{baes03}, $(\mu,\nu)=(0.11,1.27)$ (see also \citet{ferrarese02}). 
%Finally, there are three undetermined parameters ($\gamma$,$M_\mathrm{DMH}(r_\mathrm{vir})$,$\eta$) for the slowly accelerating outflows. 
%%2015/3/17
%We adopted parameters by \citet{baes03}, $(\mu,\nu)=(0.11,1.27)$ (see also \citet{ferrarese02}). 
%As a result, there are three parameters, ($\gamma$,$M_\mathrm{DMH}(r_\mathrm{vir})$,$\eta$) for slow-accelerating outflows. 

For high-velocity outflows in star-forming galaxies, we assume that the sum of the thermal and kinetic energies in the inner region is significantly larger than the gravitational potential energy, because efficient star formation supplies an extremely large amount of thermal energy to the inner star-forming region.
Thus, we can estimate the supplied $E$ from Eq. (\ref{eq_energy}) as
%%2015/7/22
%For high-velocity outflows in star-forming galaxies, we assume that sum of the thermal energy and the kinetic energy in the inner region is much larger than the gravitational potential energy, because efficient star formation supplies huge amount of thermal energy in the inner star-forming region. 
%Thus, we can estimate the supplied energy $E$ from Eq. (\ref{eq_energy}) as
%%2015/3/17
%For high-velocity outflows in starbursts, we assume that thermal energy and kinetic energy in the inner region is larger than gravitational potential energy because the large star formation supplies huge energy in the inner star-forming region. 
%Thus, we can estimate the supplied energy $E$ from Eq. (\ref{eq_energy}), 

\begin{align}
E &= \left.\left(\frac{c_\mathrm{s}^2}{\gamma-1}+\frac{v^2}{2}\right)\right|_\mathrm{r=r_\mathrm{0}} + \Phi(r_\mathrm{0}), \\
&= \eta' \left|\left(\left.\int\frac{GM_\mathrm{DMH}(r_\mathrm{vir})}{r^2}dr\right|_\mathrm{r=r_\mathrm{0}} - \frac{GM_\mathrm{BH}}{r_\mathrm{0}}\right)\right|, \nonumber\\
&\qquad + \left.\int\frac{GM_\mathrm{DMH}(r_\mathrm{vir})}{r^2}dr\right|_\mathrm{r=r_\mathrm{0}} - \frac{GM_\mathrm{BH}}{r_\mathrm{0}}, \\
&= (\eta'-1) \left|\left(\left.\int\frac{GM_\mathrm{DMH}(r_\mathrm{vir})}{r^2}dr\right|_\mathrm{r=r_\mathrm{0}} - \frac{GM_\mathrm{BH}}{r_\mathrm{0}}\right)\right|, \label{eq_burst}
\end{align}
where $r_\mathrm{0}$ and $\eta'$ are the specific inner radius and correction parameter, respectively.
Note that $r_\mathrm{0}$ represents the typical radius of the star formation region.
We assume that the star-forming region extends within the scale length of the DMH, and we set $r_\mathrm{0}=0.1r_\mathrm{d}$.
The coefficient $0.1$ is close to the observed value; the ratio of half-light radius and $r_\mathrm{d}$ in the Sombrero Galaxy is $6.1/36.1\approx0.17$ and that of the Milky Way is $6.0/38.6\approx0.16$ \citep{bendo06,sakamoto03}.
Further, $\eta'$ represents the difference between the gravitational potential energy and the sum of the thermal and kinetic energies.
Because the sum of the thermal energy and the kinetic energy is dominant over the gravitational potential energy in high-velocity outflows, $\eta'$ is sufficiently larger than unity.
Thus, there are also three parameters ($\gamma$, $M_\mathrm{DMH}(r_\mathrm{vir})$, $\eta'$) for high-velocity outflows.

We show the actual parameter ranges in Figs. \ref{fig3} and \ref{fig4}.
The parameter ranges of the quiescent galaxies indicated by Eq. (\ref{eq_slow}) are distributed within a narrow region (the black hatched region in Figs. \ref{fig3} and \ref{fig4}).
For small $\gamma$ close to unity (quasi-isothermal state), the solutions become B-1 type.
An outer transonic point is located in the outer region (within $\sim$ several hundred kiloparsec) in these types of solutions.
Thus, in the quasi-isothermal case (small $\gamma$), it is possible for the slowly accelerating outflows to have a wide subsonic region.
Also, type-B-1 solutions have an $X_\mathrm{in}$ solution.
As these two transonic solutions have different mass fluxes and starting points, the transonic solutions have different effects on galactic evolution and the release of metals into the intergalactic medium.
For example, when $\eta$ is large with small $\gamma$, the driving energy $E$ becomes large and the outer transonic point moves to the inner region (see Fig.\ref{fig6}).
For small $\eta$, which indicates a small energy supply, the solutions change from A-1 to B-1 type.
For large $\gamma$, the outflow solutions become A-1 type, with a single transonic point in the inner region ($\sim$ 0.01 kpc).
Thus, slowly accelerating outflows do not exist in cases with large $\gamma$.
Finally, when $\eta$ becomes large with large $\gamma$, the locus of the single transonic point in the type A-1 solutions does not change dramatically (see Fig.\ref{fig6}).

We also find that the solution types are not strongly dependent on the DMH mass.
This result is the same as that given by the isothermal model proposed by \citet{igarashi14}.
The Sombrero Galaxy (the trace of the subsonic outflow has been observed by Li et al. 2011) has parameters indicating that it is of type B-1.
The observed gas density distribution in this galaxy is similar to the behaviour of subsonic outflows having a transonic point at a region distant from the stellar distribution.
This result is also the same as the conclusion given in \citet{igarashi14}.
%%2015/7/22
%We also find that types of solutions do not strongly depend on DMH mass. 
%This result is the same as the isothermal model proposed by \citet{igarashi14}. 
%The Sombrero galaxy (the trace of subsonic outflow is observed by \citet{li11}) has parameters indicating type B-1. 
%The observed gas density distribution in this galaxy is similar to the subsonic outflows having transonic point at distant region from the stellar distribution. 
%This result is also the same as \citet{igarashi14} concluded. 
%%2015/4/14
%We also find that types of solutions do not strongly depend on DMH mass. 
%This result is the same as the isothermal model proposed by \citet{igarashi14}. 
%The Sombrero galaxy having the trace of subsonic outflow observed by \citet{li11} has parameters indicating type B-1. 
%The observed gas density distribution in this galaxy is similar to the subsonic outflows having transonic point at far region from stellar distribution. 
%This result is also the same as \citet{igarashi14} concluded. 
%%2015/3/17
%In addition, types of solutions are not largely dependent on DMH mass. 
%This result is the same as isothermal model proposed by \citet{igarashi14}. 
%The Sombrero galaxy having the trace of subsonic outflow observed by \citet{li11} has parameters indicating type B-1. 
%The observed gas density distribution in this galaxy is similar to the subsonic outflows having transonic point at far region from stellar distribution. 
%This result is also the same as \citet{igarashi14}. 

While the parameter range of active star-forming galaxies indicated by Eq. (\ref{eq_burst}) is widely distributed (the red hatched region in Figs. \ref{fig3} and \ref{fig4}), the majority of the solutions are categorised as type A-1.
Thus, high-velocity outflows in the star-forming galaxies have a single transonic point in the inner region, where the gravitational potential of the SMBH is dominant over that of the DMH.
The locus of this transonic point moves to the inner region for large $\eta'$, indicating a large energy supply.
This result is supported by previous observations of supersonic outflows from star-forming galaxies to which a large amount of energy is supplied \citep{strickland02}.
In addition, this property is independent of $\gamma$, and the transonic points are not strongly dependent on the DMH mass.
These results indicate that outflows in active star-forming galaxies become supersonic in the inner region, because the supplied energy strongly dominates the gravitational potential energy.

\subsection{Differences between polytropic and isothermal models}

\subsubsection{Terminal velocity}

In the isothermal model, the outflow velocity increases without limit.
However, the outflow velocity in actual galaxies is believed to converge to a finite terminal velocity $v_{\infty}$ that is dependent on $E$.
We estimate this $v_{\infty}$ in the polytropic model as follows.
From Eq. (\ref{eq_mach_number}), we can obtain the velocity equation
\begin{align}
(\gamma-1)^{-\frac{2}{\gamma-1}} v^{-2} \left\{ (E-\Phi)-\frac{v^2}{2} \right\}^{-\frac{2}{\gamma-1}} = (\gamma K)^{-\frac{2}{\gamma-1}} \dot{M}^{-2} r^4.
\end{align}
When $r \rightarrow \infty$ and $\Phi(\infty)=0$, $v_{\infty}=\sqrt{2E}$ and the terminal sound speed $c_{s\infty}=0$, or $v_{\infty}=0$ and $c_{s\infty}=\sqrt{(\gamma-1)E}$.
This indicates that $v_{\infty}$ is determined by $E$ and is independent of $\dot{M}$, $\gamma$, $K$, and $\Phi$.
The value of $c_{s\infty}$ is determined by $E$ and $\gamma$.
When $E$ is negative, $v_{\infty}$ (or $c_{s\infty}$) becomes imaginary.
This indicates that the gas cannot spread out to infinity.
The range filled by the gas is dependent on $E$ and the gravitational potential.

\begin{figure*}
 \centering
 \includegraphics[width=2.\columnwidth]{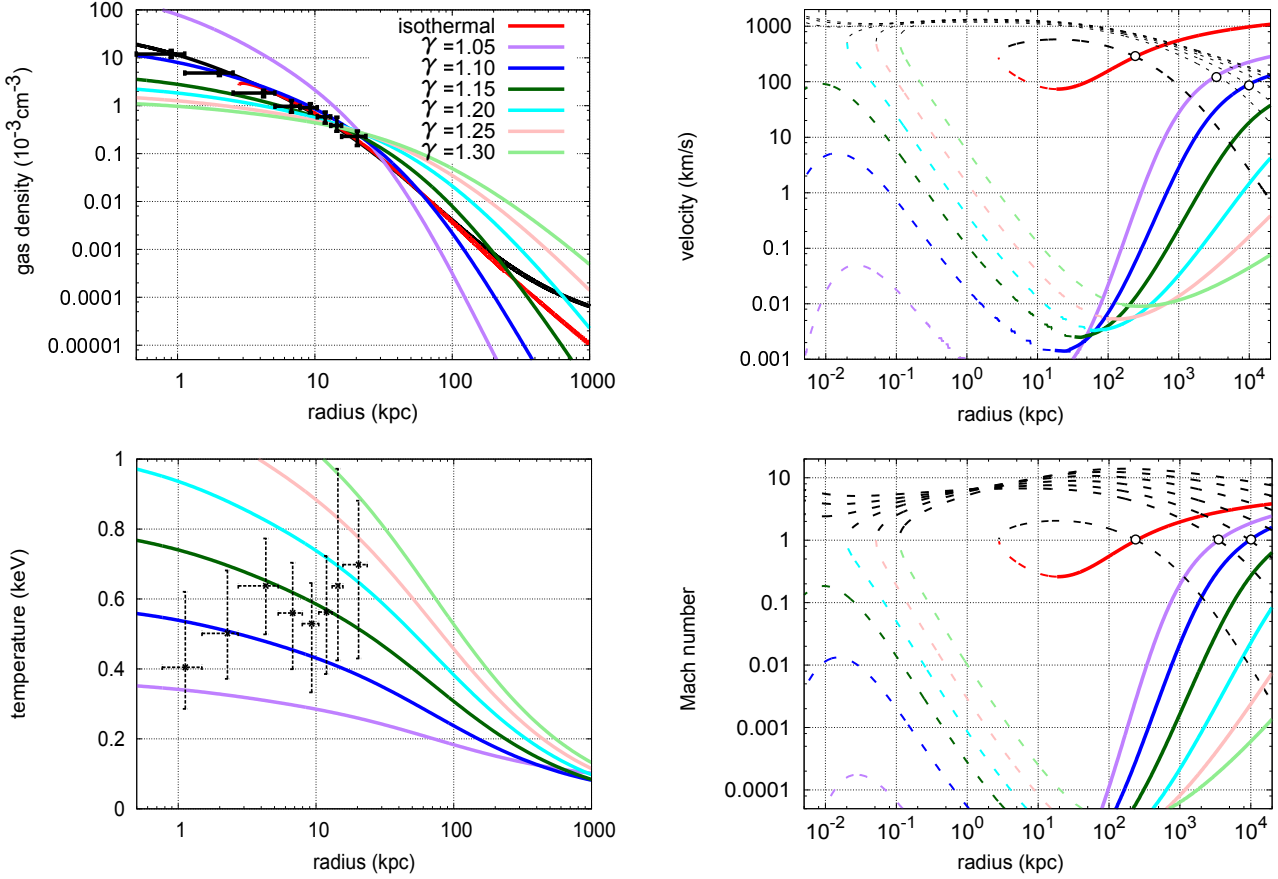}
 \caption{Transonic solutions for various $\gamma$ in the Sombrero Galaxy.
For the isothermal model, we use the result from \citet{igarashi14} at a temperature of 0.5 keV.
For the polytropic model, to determine the $E$ in Eq. (\ref{eq_energy}), the gas density is fitted to the data observed by \citet{li11} with a fixed $\dot{M} = 0.45~\mathrm{M_{\odot}} ~\mathrm{yr}^{-1}$.
The black and red lines indicate the isothermal hydrostatic and isothermal models, respectively. 
The purple, blue, dark green, cyan, pink, and light-green lines indicate the polytropic model with $(\gamma,\sqrt{E}~[\mathrm{km} ~\mathrm{s}^{-1}])=(1.05,549)$, $(\gamma,\sqrt{E})=(1.10,251)$, $(\gamma,\sqrt{E})=(1.15,117)$, $(\gamma,\sqrt{E})=(1.20,50.1)$, $(\gamma,\sqrt{E})=(1.25,19.4)$, and $(\gamma,\sqrt{E})=(1.30,6.60)$, respectively.
The solid parts of the colored lines in the right panels denote the outside of the O-point, while the dotted parts of them denote the inside of the O-point.
The white dots correspond to the positions of the transonic points.
The crosses denote observed data reported by \citet{li11}. 
}
 \label{fig7}
\end{figure*}

\subsubsection{Mass flux from Sombrero Galaxy}\label{Mass flux from Sombrero Galaxy}

\citet{igarashi14} applied the isothermal model to the Sombrero Galaxy and estimated the mass flux.
However, the fitted mass flux of $1.7$--$8.7~\mathrm{M_{\odot}} ~\mathrm{yr}^{-1}$ proposed in that study is unrealistically large compared to the expected supplied gas mass of $0.4$--$0.5~\mathrm{M_{\odot}} ~\mathrm{yr}^{-1}$, from SNe and stellar winds.
This result obviously conflicts with the assumption of stationarity in that model as,
if steady galactic winds exist in this galaxy, the mass flux must be balanced with the supplied gas mass from stars.
Here, we adopt the proposed polytropic model to resolve this problem.
We intend to fit the mass flux to the observed gas density \citep{li11} with a fixed mass flux of $0.45~\mathrm{M_{\odot}}~\mathrm{yr}^{-1}$.
We use the gravitational potential of the DMH, SMBH, and the stars, similar to the approach used to obtain Fig. 7 of \citet{igarashi14}.
Because the observed temperature distribution is close to that of the isothermal state, we focus on $\gamma\sim1$ (corresponding to an assumption of isothermality), but we estimate transonic outflows also for large $\gamma$.

As $\gamma$ increases from 1, the $E$ indicated by the fitting decreases and the transonic points move outward.
The transonic solutions with small $\gamma$ become B-2 type and the $\rmn{X_{out}}$ solution originates at the centre, while the solutions with large $\gamma$ become B-1 type and the $\rmn{X_{out}}$ solution does not originate at the centre.
The fitted gas density with $\gamma<1.10$ has a steeper slope than the observed density, while that with $\gamma>1.10$ has a shallower slope than the observed result.
Hence, we find that the polytropic transonic solution with $(\gamma,\dot{M}~[\mathrm{M_{\odot}}~\mathrm{yr}^{-1}],\sqrt{E}~[\mathrm{km} ~\mathrm{s}^{-1}])=(1.10,0.45,251)$ can reproduce the observed gas density well.
Therefore, we can conclude that the polytropic model improves upon the result provided by the isothermal model well.
Additionally, although we do not use the observed temperature data to determine $E$, the temperature distribution shown in Fig. \ref{fig7} indicates that the polytropic model can roughly reproduce the observed temperature data with $\gamma=1.10$.
This result also indicates that the polytropic model can reproduce observations well.
Although the Sombrero Galaxy is a quiescent galaxy, the possible existence of a slowly accelerating outflow in this galaxy is also shown, as in \citet{igarashi14}.
Thus, the outflow from this quiescent galaxy differs from that considered in the majority of the previous studies focusing on outflows from active star-forming galaxies.

The velocity distribution values of the polytropic model are lower than those of the isothermal model, because of the limited energy considered in the former in contrast to the unlimited energy in the latter.
We propose that this low velocity in the inner region (within $\sim10$ kpc) is due to the absence of energy and mass injection along the flow lines from the stellar components.
Thus, we expect that the slowly accelerating outflow increases the speed at the outside of the visible scale of the galaxy because the flow velocity efficiently increases after the locus of the O-point that is close to the edge of the stellar distribution (the half-light radius $\sim$ 6.1kpc).
Because mass and  energy injection decelerate and accelerate outflows, respectively, the balance between energy and mass injection along the flow lines is important if a plausible velocity distribution in the stellar region (within several kpc from the centre) is to be reproduced.
Therefore, we will investigate outflows with energy and mass injection along the flow lines in future work.
In addition, we will estimate the velocity distributions of other galaxies using the novel polytropic model proposed in this paper.

\subsubsection{Velocity distributions in actual galaxies}\label{Velocity distributions in actual galaxies}

Next, we focus on velocity distributions in actual galaxies and clarify the influence of $\gamma$.
We assume that the DMH mass is $10^{12}\mathrm{M_{\odot}}$ and use the gravitational potential indicated in Section \ref{parameters}.
As a result, the SMBH mass, $r_\mathrm{vir}$, and $r_\mathrm{d}$ become approximately $10^7\mathrm{M_{\odot}}$, $258$ kpc, and $25$ kpc, respectively.
%%2015/8/3
%Next, we focus on velocity distributions in physical galaxies and clarify the influence of $\gamma$.
%We assume that the DMH mass is $10^{12}\mathrm{M_{\odot}}$ and use the gravitational potential indicated in Section \ref{parameters}.
%As a result, the SMBH mass, $r_\mathrm{vir}$, and $r_\mathrm{d}$ become approximately $10^7\mathrm{M_{\odot}}$, $258$ kpc, and $25$ kpc, respectively.
%%2015/7/22
%Next, we focus on the velocity distribution in the actual galaxy and clarify the influence of polytropic index $\gamma$. 
%We assume that the DMH mass is $10^{12}\mathrm{M_{\odot}}$ and use the gravitational potential indicated in Section \ref{parameters}. 
%As a result, the SMBH mass and the virial radius and the scale radius of DMH become approximately $10^7\mathrm{M_{\odot}}$, $258$kpc and $25$kpc, respectively. 

To determine the $E$ in Eq. (\ref{eq_energy}), we assume that the sum of the thermal and kinetic energies at the starting point can be approximated to the supplied energy per unit mass from SNe.
The supplied energy per unit mass is the ratio of the energy injection per unit time $\dot{e}$ to the mass injection per unit time $\dot{M}$ (the mass flux).
Under these assumptions, $E$ becomes
\begin{align}
E = \frac{\dot{e}}{\dot{M}} + \Phi(r_\mathrm{s}), \label{eq_energy_sombrero}
\end{align}
where $r_\mathrm{s}$ is the radius of the starting point of the outflow.
If the mass and energy supply are determined by the stellar distribution, the starting point is approximately several times the scale radius of the stellar distribution (typically several kpc in a $10^{12}\mathrm{M_{\odot}}$ galaxy).
Hence, we assume $r_\mathrm{s}=5$ kpc.

We also define 
\begin{align}
& \dot{e} = \lambda f_\mathrm{SN} \left(\frac{\dot{M}_\mathrm{\ast}}{\mathrm{M_{\odot}}}\right) \times 10^\mathrm{51}\mathrm{erg}, \\
& \dot{M} = R_\mathrm{f} \dot{M}_\mathrm{\ast},
\end{align}
where $\dot{M}_\mathrm{\ast}$, $\lambda$, and $f_\mathrm{SN}$ are the star formation rate, the fraction of energy retained after the radiative energy losses, and the energy injection rate from the SNe, respectively.
The factor $R_\mathrm{f}$ is the return fraction to the interstellar medium.
In the Kroupa-Chevalier initial mass function, $R_\mathrm{f}=0.257$ and $f_\mathrm{SN}=1.86\times10^{-2}$.
Additionally, we assume $\dot{M}_\mathrm{\ast}=10~\mathrm{\mathrm{M_{\odot}}}~\mathrm{yr}^{-1}$ since the star formation rate in starburst galaxies is typically several tens of solar mass per year.

As a result, the $E$ per unit mass in Eq. (\ref{eq_energy_sombrero}) and $\dot{M}$ become
\begin{align}
& E = 7.23 \times 10^{49} ~[\mathrm{erg} ~\mathrm{M_{\odot}^{-1}}] \times \lambda  \left(\frac{f_\mathrm{SN}}{1.86\times10^{-2}}\right) \left(\frac{R_\mathrm{f}}{0.257}\right)^{-1}, \nonumber\\
&\quad\quad\quad\quad\quad  +\Phi(r_\mathrm{s}), \\ \label{eq_energy_sombrero_2}
& \dot{M} = 2.57 ~[\mathrm{\mathrm{M_{\odot}}}~\mathrm{yr}^{-1}] \times \left(\frac{R_\mathrm{f}}{0.257}\right) \left( \frac{\dot{M}_\mathrm{\ast}}{10} \right).
\end{align}

\begin{figure}
 \centering
 \includegraphics[width=1\columnwidth]{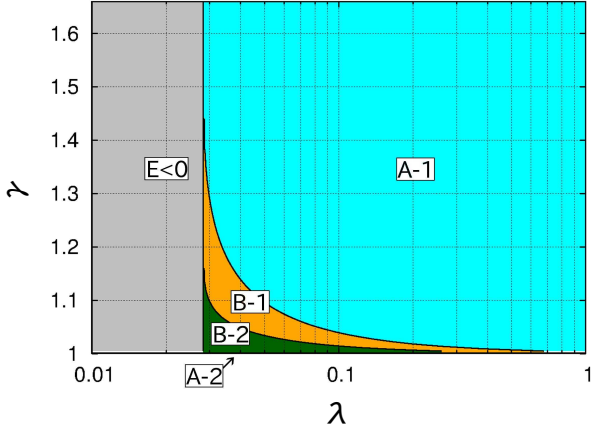}
 \caption{Various model solutions incorporating DMH and SMBH gravitational potentials in an actual galaxy with $M_\mathrm{DMH}(r_\mathrm{virial}) = 10^{12}~\mathrm{M_{\odot}}$.
The horizontal and vertical axes represent $\lambda$ and $\gamma$, respectively.
The colours of the various regions correspond to the solution types shown in Fig. \ref{fig3}.
Transonic solutions do not exist in the grey region, because the specific energy is negative. }
 \label{fig8}
\end{figure}

We show the various transonic solutions in Fig. \ref{fig8}.
When $\gamma$ and $\lambda$ are large, there is only one transonic solution (type A-1).
Thus, in this parameter region, the transonic outflows originate at the centre and the transonic point is in the inner region ($\sim$ 0.01 kpc).
For small $\gamma$ and large $\lambda$, there are two transonic solutions.
The outer transonic point is at the far distance (within several hundreds of kiloparsec) and the transonic solution through this point has a wide subsonic region.
This widely spread subsonic region is similarly indicated by the isothermal model \citep{igarashi14}.
When $\lambda$ is small, $E$ becomes negative and there is no transonic solution.

\begin{figure*}
 \centering
 \includegraphics[width=2.\columnwidth]{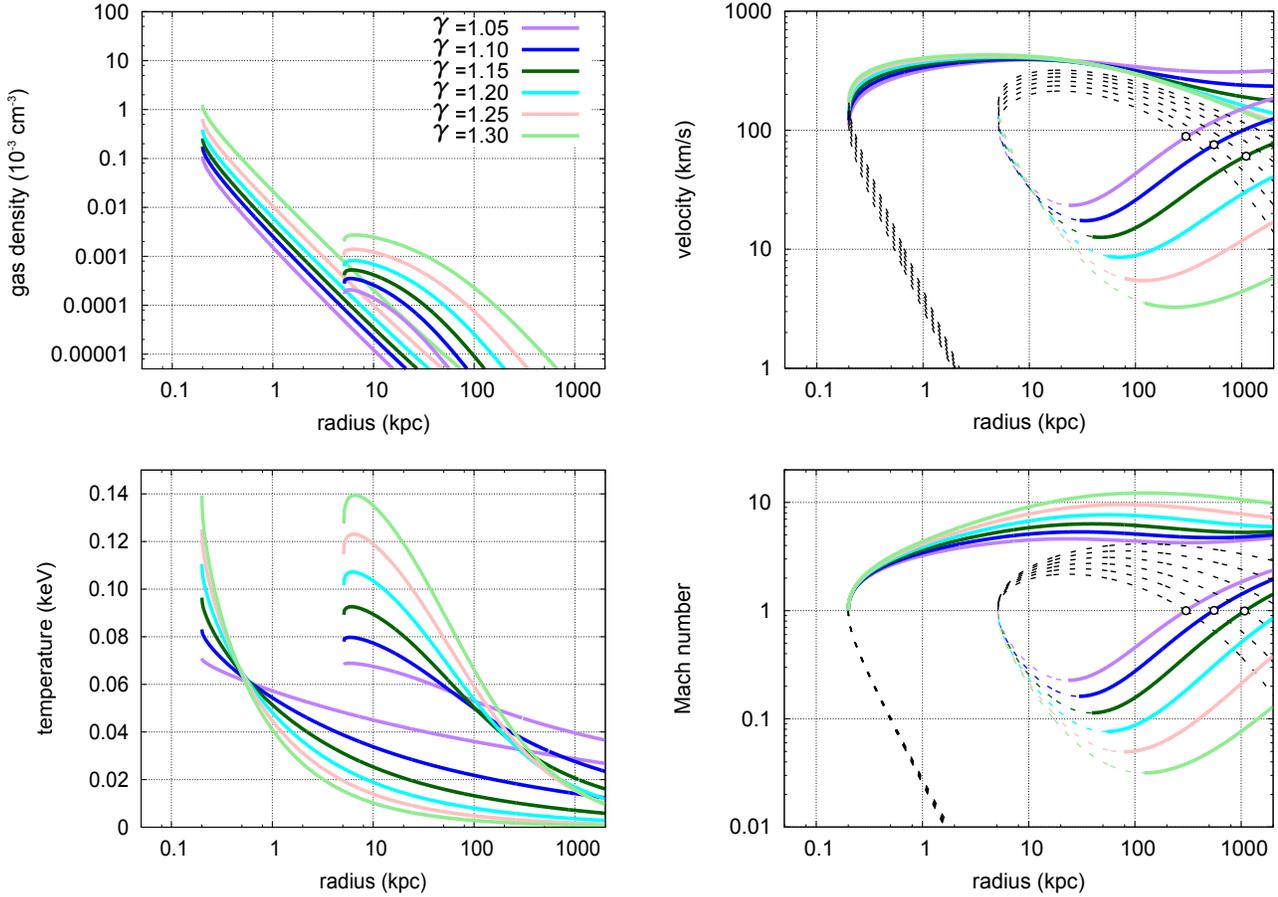}
 \caption{Transonic solutions for various $\gamma$ in an actual galaxy with $M_\mathrm{DMH}(r_\mathrm{virial}) = 10^{12}\mathrm{M_{\odot}}$.
To determine the $\lambda$ in Eq. (\ref{eq_energy_sombrero_2}), the starting point is fit to $r_\mathrm{start}=5$ kpc.
The purple, blue, dark green, cyan, pink, and light-green lines indicate polytropic model results with $(\gamma,\lambda)=(1.05,0.0650)$, $(\gamma,\lambda)=(1.10,0.0399)$, $(\gamma,\lambda)=(1.15,0.0325)$, $(\gamma,\lambda)=(1.20,0.0297)$, $(\gamma,\lambda)=(1.25,0.0285)$, and $(\gamma,\lambda)=(1.30,0.0281)$, respectively.
Solutions having starting points at 200 pc are supersonic solutions with the same parameters as the transonic solutions. 
The solid parts of the colored lines in the right panels denote the outside of the O-point, while the dotted parts of them denote the inside of the O-point.
The white dots correspond to the positions of the transonic points.
}
 \label{fig9}
\end{figure*}

To clarify the influence of $\gamma$ on the polytropic model, we alter $\gamma$ arbitrarily and estimate $\lambda$ by fitting starting points for the assumed starting point of $r_\mathrm{s}=5$ kpc.
The results are shown in Fig. \ref{fig9}.
When $\gamma$ is close to 1 (corresponding to an isothermal state), the fitted $\lambda$ is large and the transonic point is within several hundreds of kiloparsec.
Small $\gamma$ indicates that a large amount of energy is contained in the interstellar medium.
This is consistent with the fact that large $\lambda$ indicates that a large amount of energy is retained after radiative cooling.
When $\gamma$ is large, the fitted $\lambda$ becomes small and the transonic point moves outward.
Finally, small $\gamma$ with small $\lambda$ indicates that a small amount of energy is retained in the interstellar medium.
Thus, it is difficult to accelerate outflows and the outer transonic point is at the far distance.

Because previous studies have focused on supersonic solutions, we show these solutions in Fig. \ref{fig9}.
\citet{chevalier85} and \citet{sharma13} fixed starting points at $(r~[$pc$],\mathcal{M})=(200,1)$ for supersonic solutions.
Therefore, we also fix starting points in the same position.
The velocities of both the transonic and supersonic solutions converge to the same $v_{\infty}$ in the far distance, but the slopes of the velocity distributions differ considerably.
The transonic solution is accelerated beyond the O-point ($\geq 100$ kpc), while the supersonic solution is decelerated in this region.
The density and the temperature of both solutions decrease, but those of the transonic solution are higher than those of the supersonic solution.
In the subsonic region, because supersonic outflows can lead to reduced density and temperature through rapid expansion with high velocity, and because the transonic solution becomes supersonic in the far distance, the density and the temperature of the transonic solution are also higher than those of the supersonic solution.

\section{Conclusion} \label{conclusion}

We have revealed polytropic transonic solutions of spherically symmetric and steady galactic winds considering the gravitational potentials of a dark matter halo (DMH) and supermassive black hole (SMBH).
These solutions have been classified in terms of their topological features in the diagrams shown in Figs. \ref{fig1} and \ref{fig2}.
Further, we have classified the transonic solutions as shown in Figs. \ref{fig3} and \ref{fig4}.
Similar to \citet{igarashi14}, we conclude that the gravitational potential of the SMBH adds a new branch to the transonic solutions generated by considering the gravity of the DMH.
The inner transonic point is formed by the gravity of the SMBH, whereas the outer transonic point is due to that of DMH.
The transonic solution types depend on the mass distribution, the amount of supplied energy, the polytropic index $\gamma$, and the slope $\alpha$ of the DMH mass distribution.
When $\alpha$ becomes larger than a critical value $\alpha_\mathrm{c}$, the transonic solution types change dramatically.
We have also found the analytical relation between $\alpha_\mathrm{c}$ and $\gamma$.

We have estimated the parameter ranges for actual galaxies using the results of prior studies.
The most interesting feature revealed by this model is the possibility that two transonic points may occur in the quasi-isothermal state (small $\gamma$).
These two transonic solutions have different mass fluxes and different starting points.
Thus, the transonic solution type influences the galactic evolution and the release of metals from galaxies.
The transonic solution through the outer transonic point generated by the gravitational potential of the DMH indicates the possibility of a slowly accelerating outflow.
We have found that it is possible for this slowly accelerating outflow to exist even in quiescent galaxies with small $\gamma$.
Further, this slowly accelerating outflow differs from the results of many previous studies focusing on supersonic outflows in active star-forming galaxies.
Also, with regard to active star-forming galaxies, our model indicates that outflows have only one transonic point in the inner region ($\sim0.01$ kpc).
This result is not strongly dependent on $\gamma$. 

We have predicted that the polytropic model has a different mass flux from that of the isothermal model, because the isothermal wind can be supplied with an unlimited amount of energy, whereas the energy supplied to the polytropic wind is limited.
For example, we have applied the polytropic model with mass flux supplied by stellar components to the Sombrero Galaxy, and concluded that this approach can reproduce the observed gas density and temperature distributions well.
This result differs significantly from the isothermal model, which has unrealistically large mass flux \citep{igarashi14}.
Thus, we have concluded that the polytropic model is more realistic than the isothermal model, and that the Sombrero Galaxy has a slowly accelerating outflow.
It is difficult to observe this slowly accelerating outflow via X-ray monitoring, because the outflow velocity in this galaxy is expected to be low.
We expect that next-generation X-ray observation satellites will have the ability to detect the detailed structure of these outflows with higher resolution.
In future research, we will construct an outflow model incorporating energy and mass injection along the flow lines in order to more accurately reproduce the observed velocity distributions.

Finally, we have focused on the influence of $\gamma$ on the transonic solutions and have investigated the gas density, temperature, and velocity distributions of an actual galaxy with DMH mass $\sim10^{12}\mathrm{M_{\odot}}$.
We have concluded that the slopes of the transonic solution distributions are strongly dependent on $\gamma$ and the amount of energy in the system.
We have also compared the transonic and supersonic solutions, and found that the transonic solutions accelerate in the region beyond the O-point while the supersonic solutions decelerate.

In addition, we prove that the transonic solution is entropy-maximum independently of the structure of the mass density distributions.
In actual galaxies, the mass density distributions including DMH, SMBH and stellar mass are complicated but this proof indicates that the complexity of these gravitational sources does not influence the availability of the transonic solution.
The gravitational potentials influence only the topological feature of the transonic solution. 
%%2016/10/12
%In addition, we prove that the transonic solution is entropy-maximum independently of the structure of the mass density distributions.
%In actual galaxies, the mass density distributions including DMH, SMBH and stellar mass are complicated but this proof indicates that the complexity of these gravitational sources does not influence the validity of the transonic solution.
%The gravitational potentials influence only the topological feature of the transonic solution. 

\section*{Acknowledgements}

This work was supported in part by a JSPS Grant-in-Aid for Scientific Research: (C) (25400222, 20540242), and by a Mitsubishi Foundation Research Grant in the Natural Sciences, No. 25134. This research used computational resources in Center for Computational Sciences, University of Tsukuba.
%%2016/5/30
%This work was supported in part by a JSPS Grant-in-Aid for Scientific Research: (C) (25400222, 20540242), and by a Mitsubishi Foundation Research Grant in the Natural Sciences, No. 25134.
%%2015/7/22
%This work was supported in part by JSPS Grants-in-Aid for Scientific Research: (C) (25400222, 20540242) and the Mitsubishi Foundation Research Grants in the Natural Sciences 25134. 

\appendix

\section{Feasibility of the transonic flow} \label{A}

In this Appendix, we prove that the entropy of the transonic solution is maximum independent of concrete functional form of the spatial distribution of the mass density. 
Assuming a polytropic, steady and spherically symmetric flow, it can be proofed based on Eq. (\ref{eq_mach_number}).
We define the left hand side of Eq. (\ref{eq_mach_number}) as
\begin{align}
f(\mathcal{M}) = \mathcal{M}^{-1} \{ (\gamma-1)\mathcal{M}^2+2 \}^{\frac{\gamma+1}{2(\gamma-1)}}, \label{eq_appendix_1}
\end{align}
and the right hand side as
\begin{align}
g( &r, K,\dot{M},E) \nonumber\\
&= \{2(\gamma-1)\}^{\frac{\gamma+1}{2(\gamma-1)}} (\gamma K)^{-\frac{1}{\gamma-1}} \dot{M}^{-1} r^2 \{E-\Phi(r)\}^{\frac{\gamma+1}{2(\gamma-1)}}. \label{eq_appendix_2}
\end{align}
By differentiating Eq. (\ref{eq_appendix_1}) with respect to $\mathcal{M}$, we obtain
\begin{align}
\frac{\partial f(\mathcal{M})}{\partial \mathcal{M}} = 2\left( 1-\frac{1}{\mathcal{M}^2} \right) \{ (\gamma-1)\mathcal{M}^2+2 \}^{\frac{3-\gamma}{2(\gamma-1)}}. \label{eq_appendix_3}
\end{align}
Eq. (\ref{eq_appendix_3}) has a singularity at $\mathcal{M}=1$.
This derivative is negative for $\mathcal{M}<1$ and positive for $\mathcal{M}>1$.
%%2015/12/2
%In this Appendix, we prove that the transonic solution is entropy-maximum independently of the spacial structure of the mass density distribution.
%Assuming a polytropic, steady, spherically symmetric flow, we use Eq. (\ref{eq_mach_number}) for this proof problem.
%We define the left hand side of Eq. (\ref{eq_mach_number}) as
%\begin{align}
%f(\mathcal{M}) = \mathcal{M}^{-1} \{ (\gamma-1)\mathcal{M}^2+2 \}^{\frac{\gamma+1}{2(\gamma-1)}}, \label{eq_appendix_1}
%\end{align}
%and the right hand side of that as
%\begin{align}
%g( &r, K,\dot{M},E) \nonumber\\
%&= \{2(\gamma-1)\}^{\frac{\gamma+1}{2(\gamma-1)}} (\gamma K)^{-\frac{1}{\gamma-1}} \dot{M}^{-1} r^2 \{E-\Phi(r)\}^{\frac{\gamma+1}{2(\gamma-1)}}. \label{eq_appendix_2}
%\end{align}
%By differenciating Eq. (\ref{eq_appendix_1}) with respect to $\mathcal{M}$, we obtain
%\begin{align}
%\frac{\partial f(\mathcal{M})}{\partial \mathcal{M}} = 2\left( 1-\frac{1}{\mathcal{M}^2} \right) \{ (\gamma-1)\mathcal{M}^2+2 \}^{\frac{3-\gamma}{2(\gamma-1)}}. \label{eq_appendix_3}
%\end{align}
%Eq. (\ref{eq_appendix_3}) has a singularity at $\mathcal{M}=1$.
%This derivative is negative when $\mathcal{M}<1$, and positive when $\mathcal{M}>1$.
%Therefore, $f(\mathcal{M})$ monotonically decreases when $\mathcal{M}<1$, increases when $\mathcal{M}>1$ and becomes minimum at $\mathcal{M}=1$.

At the locus of the X-point, Eq. (\ref{eq_appendix_2}) becomes
\begin{align}
g( &r_\mathrm{X}, K,\dot{M},E) \nonumber\\
&= \{2(\gamma-1)\}^{\frac{\gamma+1}{2(\gamma-1)}} (\gamma K)^{-\frac{1}{\gamma-1}} \dot{M}^{-1} r_\mathrm{X}^2 \{E-\Phi(r_\mathrm{X})\}^{\frac{\gamma+1}{2(\gamma-1)}}, \label{eq_appendix_4}
\end{align}
noting that $r_\mathrm{X}$ is the locus of the X-point.
If we ignore the energy and the mass injection along the stream line ($E$ and $\dot{M}$ are constant), the entropy $K$ becomes maximum at $\mathcal{M}=1$ because $f(\mathcal{M})(=g(K))$ becomes minimum at $\mathcal{M}=1$.
We can conclude that the entropy of the subsonic solution ($\mathcal{M}<1$ everywhere) and the supersonic solution ($\mathcal{M}>1$ everywhere) is smaller than that of the transonic solution.
This result make it clear that the entropy of the transonic flow is maximum among the solutions connecting the starting point of the flow and the infinity.
We must note that this result is universal and independent of the functional form of the gravitational potential.
In actual galaxies, the galactic mass density distribution is complicated but the galactic outflow should become the transonic solution with the maximum entropy.
%2015/12/2
%At the locus of the X-point, Eq. (\ref{eq_appendix_2}) becomes
%\begin{align}
%g( &r_\mathrm{X}, K,\dot{M},E) \nonumber\\
%&= \{2(\gamma-1)\}^{\frac{\gamma+1}{2(\gamma-1)}} (\gamma K)^{-\frac{1}{\gamma-1}} \dot{M}^{-1} r_\mathrm{X}^2 \{E-\Phi(r_\mathrm{X})\}^{\frac{\gamma+1}{2(\gamma-1)}}. \label{eq_appendix_4}
%\end{align}
%noting $r_\mathrm{X}$ is the locus of the X-point.
%When $E$ and $\dot{M}$ are constant, $K$ becomes largest at $\mathcal{M}=1$ because $f(\mathcal{M})(=g(K))$ becomes minimum at $\mathcal{M}=1$.
%This indicates that the entropy of the transonic flow is maximum comparing to that of the subsonic flow ($\mathcal{M}<1$) and the supersonic flow ($\mathcal{M}>1$).
%This result is independent on the form of the graviatational potential.
%In actual galaxies, the galactic mass density distribution is complicated but the galactic outflow should become the transonic solution with the maximum entropy.

Similarly to this discussion, we can also prove that the massflux/the energy of the transonic solution are maximum/minimum as follows.
If we ignore the energy injection and the entropy production along the stream line ($E$ and $K$ are constant), the massflux $\dot{M}$ becomes maximum at $\mathcal{M}=1$.
Therefore, the massflux of the transonic flow is maximum among the solutions connecting the starting point of the flow and the infinity.
Moreover, if we ignore the mass injection and the entropy production along the stream line ($\dot{M}$ and $K$ are constant), the energy $E$ becomes minimum at $\mathcal{M}=1$.
Therefore, the energy flux of the transonic flow is minimum among the solutions connecting the starting point of the flow and the infinity.
All these results clearly show that the transonic solution is the unique and natural solution for realistic astrophysical outflows not only as the model for the galactic winds but as the model for general polytropic, steady and spherically symmetric flows.
In addition, these results are independent of the sign of the energy $E$.
Therefore, the entropy of the transonic solution is maximum if the solutions do not extend to the infinity. 
%%2015/12/2
%Similary to this proof, we can prove that the transonic solution is the massflux-maximum and energy-minimum solution.
%When $E$ and $K$ are constant, $\dot{M}$ becomes largest at $\mathcal{M}=1$.
%Thus, the massflux of the transonic flow is maximum comparing to that of the subsonic flow and the supersonic flow.
%Thus, the transonic solution is massflux-maximum.
%Moreover,when $\dot{M}$ and $K$ are constant, $E$ becomes smallest at $\mathcal{M}=1$.
%Thus, the energy of the transonic flow is minimum comparing to that of the subsonic flow and the supersonic flow.
%Thus, the transonic solution is also energy-minimum.

\bsp

\label{lastpage}


\begin{thebibliography}{99}
%\bibitem[\protect\citeauthoryear{}{}]{}
\bibitem[\protect\citeauthoryear{Adelberger et al.}{2003}]{adelberger03} Adelberger K. L., Steidel C. C., Shapley A. E., Pettini M., 2003, ApJ, 584, 45
\bibitem[\protect\citeauthoryear{Aguirre et al.}{2001}]{aguirre01} Aguirre, A., Hernquist, L., Schaye, J., Weinberg, D.H., Katz, N., \& Gardner, J. 2001, ApJ, 560, 599
\bibitem[\protect\citeauthoryear{Athey et al.}{2002}]{athey02} Athey, A., Bregman, Joel, Bregman, Jesse, Tami, P., \& Sauvage, M. 2002,
ApJ, 571, 272
\bibitem[\protect\citeauthoryear{Babul \& Rees}{1992}]{babul92} Babul, A., \& Rees, M.J. 1992, MNRAS, 255, 346
\bibitem[\protect\citeauthoryear{Baes et al.}{2003}]{baes03} Baes, M., Buyle, P., Hau, G.K. \& Dejonghe, H., 2003, MNRAS, 341, L44
\bibitem[\protect\citeauthoryear{Bajaja et al.}{1984}]{bajaja84} Bajaja, E., van der Burg, G., Faber, S.M., Gallagher, J.S., Knapp, G.R., \& Shane, W.W. 1984, A\&A, 141, 309
\bibitem[\protect\citeauthoryear{Bajaja et al.}{1991}]{bajaja91} Bajaja E., Krause, M., Dettmar, R.-J., \& Wielebinski, R. 1991, A\&A, 241, 411
\bibitem[\protect\citeauthoryear{Barnes \& Fluke}{2008}]{fluke08} Barnes, D.G., \& Fluke, C.J. 2008, New Astronomy, 13, 599
\bibitem[\protect\citeauthoryear{Behroozi, Conroy \& Wechsler}{2010}]{behroozi10} Behroozi, P.S., Conroy, C., \& Wechsler, R.H. 2010, ApJ, 717, 379
\bibitem[\protect\citeauthoryear{Bell et al.}{2003}]{bell03} Bell, E.F., McIntosh, D.H., Katz, N., \& Weinberg, M.D. 2003, ApJS, 149, 289
\bibitem[\protect\citeauthoryear{Bendo et al.}{2006}]{bendo06} Bendo G.J., Buckalew, D.A.D., Draine, B.T., Joseph, R.D., Kennicutt, R.C., Sheth, J.K., Smith, J.D.T., Walter, F., Calzetti, J.M., Engelbracht, C.W., Gorden, K.D., Helou, G., Hollenbach, D., Murphy, E.J., \& Roussel, H. 2006, ApJ, 645, 134
\bibitem[\protect\citeauthoryear{Bensby et al.}{2005}]{bensby05} Bensby, T., Feltzing, S., Lundstr\"{o}m, I., \& Ilypin, I. 2005, A\&A, 433, 185
\bibitem[\protect\citeauthoryear{Binney}{2004}]{binney04} Binney, J. 2004, MNRAS, 347, 1093
\bibitem[\protect\citeauthoryear{Binney et al.}{2009}]{binney09} Binney J., Nipoti C., Fraternali F., 2009, MNRAS, 397, 1804
\bibitem[\protect\citeauthoryear{Blumenthal et al.}{1984}]{blumenthal84} Blumenthal, G. R., Faber, S. M., Primack, J. R. \& Rees, M. J. 1984, Nature 311, 517
\bibitem[\protect\citeauthoryear{Bogd\'an \& Gilfanov}{2008}]{bogdan08} Bogd\'an, \'A., \& Gilfanov, M. 2008, MNRAS, 388, 56
\bibitem[\protect\citeauthoryear{Bogd\'an et al.}{2012}]{bogdan12} Bogd\'an, \'A., David, L.P., Jones, C., Forman, W.R., \& Kraft, R.P. 2012, ApJ, 758, 65
\bibitem[\protect\citeauthoryear{Bregman}{1980}]{bregman80} Bregman J. N., 1980, ApJ, 236, 577
\bibitem[\protect\citeauthoryear{Breitschwerdt et al.}{1991}]{breitschwerdt91} Breitschwerdt D., McKenzie J. F., V\''{o}lk H. J., 1991, A\&A, 245, 79
\bibitem[\protect\citeauthoryear{Bridges et al.}{2007}]{bridges07} Bridges, T.J., Rhode, K.L., Zepf, S.E., \& Freeman, K.C. 2007, ApJ, 658, 980
\bibitem[\protect\citeauthoryear{Bullock et al.}{2001}]{bullock01} Bllock, J.S., Kolatt, T.S., Sigad, Y., Somerville, R.S., Kravtsov, A.V., Klypin, A.A., Primack, J.R. \& Dekel, A., 2001, MNRAS, 321, 559
\bibitem[\protect\citeauthoryear{Burbidge et al.}{1964}]{burbidge64} Burbidge E. M., Burbidge G. R., Rubin V. C., 1964, ApJ, 140, 942
\bibitem[\protect\citeauthoryear{Burke}{1968}]{burke68} Burke, J.A., 1968, MNRAS, 140, 241
\bibitem[\protect\citeauthoryear{Burkert}{1995}]{burkert95} Burkert A. 1995, ApJ, 447, L25
\bibitem[\protect\citeauthoryear{Cappellaro et al.}{1999}]{cappellaro99} Cppellaro, E., Evans, R., \& Turatto, M. 1999, A\&A, 351, 459
\bibitem[\protect\citeauthoryear{Cattaneo et al.}{2006}]{cattaneo06} Cattaneo, A., Dekel, A., Devriendt, J., Guiderdoni, B., \& Blaizot, J. 2006, MNRAS, 370, 1651
\bibitem[\protect\citeauthoryear{Cen \& Ostriker}{1992}]{cen92} Cen, R. \& Ostriker, J., 1992, ApJ, 393, 22
\bibitem[\protect\citeauthoryear{Chakrabarti}{1990}]{chakrabarti90} Chakrabarti, S. K. Theory of Transonic Astrophysical Flows, 1990, World Scientific, Singapore
\bibitem[\protect\citeauthoryear{Chevalier \& Clegg}{1985}]{chevalier85} Chevalier, R.A., \& Clegg, A.W. 1985, Nature, 317, 44
\bibitem[\protect\citeauthoryear{Cole}{1991}]{cole91} Cole, S. 1991, ApJ, 367, 45
\bibitem[\protect\citeauthoryear{Davis et al.}{1985}]{davis85} Davis, M., Efstathiou, G., Frenk, C., \& White, S. D. M. 1985, ApJ, 292, 371
\bibitem[\protect\citeauthoryear{Dekel \& Silk}{1986}]{dekel86} Dekel,A., \& Silk, J. 1986, ApJ, 303, 39
\bibitem[\protect\citeauthoryear{Dekel \& Rees}{1987}]{dekel87} Dekel, A., \& Rees, M. J. 1987, Nature, 326, 455
\bibitem[\protect\citeauthoryear{Diehl \& Statler}{2008}]{diehl08} Diehl, S., \& Statler, T. S. 2008, ApJ, 687, 986
\bibitem[\protect\citeauthoryear{Efstathiou}{1992}]{efstathiou92} Efstathiou, G. 1992, MNRAS, 256, 43
\bibitem[\protect\citeauthoryear{Ellison et al.}{2000}]{ellison00} Ellison S.L., Songaila A., Schaye J., \& Pettini M. 2000, AJ, 120, 1175
%%\bibitem[\protect\citeauthoryear{Ferrarese \& Merritt}{2000}]{ferrarese00} Ferrarese, L. \& Merritt, D., 2000, ApJ, 539, L9
\bibitem[\protect\citeauthoryear{Evrard}{1988}]{evrard88} Evrard, A. E., 1988, MNRAS, 235, 911
\bibitem[\protect\citeauthoryear{Faber \& Gallagher}{1976}]{faber76} Faber, S. M. \& Gallagher, J. S., 1976, ApJ, 204, 365
\bibitem[\protect\citeauthoryear{Ferrarese}{2002}]{ferrarese02} Ferrarese, L., 2002, ApJ, 578, 90 
\bibitem[\protect\citeauthoryear{Fluke \& Barnes}{2008}]{barnes08} Fluke, C.J., \& Barnes, D.G. 2008, Astronomy Education Review, 7, 113
\bibitem[\protect\citeauthoryear{Frenk}{1991}]{frenk91} Frenk, C. S., 1991, Physica Scripta, 36, 70
\bibitem[\protect\citeauthoryear{Fukushige \& Makino}{1997}]{fukushige97} Fukushige T., \& Makino J. 1997, ApJ, 477, L9
\bibitem[\protect\citeauthoryear{Fukuzawa et al.}{2006}]{fukuzawa06} Fukuzawa, Y., Betoya-Nonesa, J.G., Pu, J., Ohto, A., \& Kawano, N. 2006, ApJ, 636, 698
%%\bibitem[\protect\citeauthoryear{G\"{u}ltekin et al.}{2009}]{gultekin09} G\"{u}ltekin, K.,Richstone, D.O., Gebhardt, K., et al. 2009, ApJ, 698, 198
%%\bibitem[\protect\citeauthoryear{Gebhardt et al.}{2000}]{gebhardt00} Gebhardt, K., Bender, R., Bower, G., et al. 2000, ApJL, 539, 13
\bibitem[\protect\citeauthoryear{Gisler}{1976}]{gisler76} Gisler, G. R., 1976, A\& A, 51, 137
%%\bibitem[\protect\citeauthoryear{Graham et al.}{2011}]{graham11} Graham, A.W., Onken, C.A. Athanassoula, E. \& Combes, F., 2011, MNRAS, 412, 2211
\bibitem[\protect\citeauthoryear{Gunn \& Gott}{1972}]{gunn72} Gunn, J. E. \& Gott, J. R. III, 1972, ApJ, 176, 1
\bibitem[\protect\citeauthoryear{Hameed \& Devereux}{2005}]{hameed05} Hameed, S., \& Devereux, N. 2005, AJ, 129, 2597
\bibitem[\protect\citeauthoryear{Hayashi \& Chiba}{2012}]{hayashi12} Hayashi, K., \& Chiba, M. 2012, ApJ, 755, 145
\bibitem[\protect\citeauthoryear{Heckman}{1980}]{heckman80} Heckman, T.M. 1980, A\&A, 87, 152
\bibitem[\protect\citeauthoryear{Heckman}{2003}]{heckman03} Heckman, T.M. 2003, Revista Mexicana de Astronomia y Astrofisica Conference Series, 17, 47
\bibitem[\protect\citeauthoryear{Hernquist}{1990}]{hernquist90} Hernquist, L., 1990, ApJ, 356, 359
\bibitem[\protect\citeauthoryear{Hernquist \& Katz}{1989}]{hernquist89} Hernquist, L. \& Katz, N., 1989, ApJS, 70, 419
\bibitem[\protect\citeauthoryear{Hessen et al.}{2009}]{hessen09} Hessen, V., Beck, R., Krause, M., \& Dettmar, R.-J. 2009, A\&A, 494, 563
\bibitem[\protect\citeauthoryear{Hopkins et al.}{2012}]{hopkins12} Hopkins, P.F., Quataert, E., Murray, N. 2012, MNRAS, 421, 3522
\bibitem[\protect\citeauthoryear{Igarashi et al.}{2014}]{igarashi14} Igarashi, A., Mori, M., \& Nitta, S. 2014, MNRAS, 444, 1177
\bibitem[\protect\citeauthoryear{Ipavich}{1975}]{ipavich75} Ipavich F. M., 1975, ApJ, 196, 107
%%\bibitem[\protect\citeauthoryear{Hu}{2008}]{hu08} Hu, J., 2008, MNRAS, 386, 2242
\bibitem[\protect\citeauthoryear{Ishiyama et al.}{2013}]{ishiyama13} Ishiyama, T. et al., 2013, ApJ, 767, 146
\bibitem[\protect\citeauthoryear{Jing \& Suto}{2000}]{jing00} Jing, Y.P.., \& Suto, Y. 2000, ApJ, 529, L69
\bibitem[\protect\citeauthoryear{Jing \& Suto}{2002}]{jing02} Jing, Y.P.., \& Suto, Y. 2000, ApJ, 574, 538
\bibitem[\protect\citeauthoryear{Johnson \& Axford}{1971}]{johnson71} Johnson, H. E., \& Axford, W. I. 1971, ApJ, 165, 381
\bibitem[\protect\citeauthoryear{Kauffmann, White \& Guiderdoni}{1993}]{kauffmann93} Kauffmann, G., White, S. D. M. \& Guiderdoni, B., 1993, MNRAS, 264, 201
\bibitem[\protect\citeauthoryear{Kennicutt}{1998}]{kennicutt98a} Kennicutt, Jr., R.C., 1998, ApJ, 498, 541
\bibitem[\protect\citeauthoryear{Kennicutt}{1998}]{kennicutt98} Kennicutt, R. C. Jr., 1998, ARA\&A, 36, 189
\bibitem[\protect\citeauthoryear{Kent}{1988}]{kent88} Kent, S.M. 1988, AJ, 96, 514
\bibitem[\protect\citeauthoryear{Klypin et al.}{2011}]{klypin11} Klypin, A., Trujillo-Gomez, S. \& Primack, J., 2011, ApJ, 740, 102
\bibitem[\protect\citeauthoryear{Knapp et al.}{1992}]{knapp92} Knapp, G.R., Gunn, J.E., \& Wynn-Williams, C.G. 1992, ApJ, 399, 76
\bibitem[\protect\citeauthoryear{Kormendy et al.}{1996}]{kormendy96} Kormendy, J., Bender, R., Ajhar, E.A., Dressler, A., Faber, S.M., Gebhardt, K., Grillmair, C., Lauer, T.R., Richstone, D., \& Tremaine, S. 1996, ApJ, 473, L91
\bibitem[\protect\citeauthoryear{Lamers \& Cassinelli}{1999}]{lamers99} Lamers, H. J. G. L. M.  \& Cassinelli, J. P. 1999, Introduction to Stellar Winds, 
Cambridge Univ. Press, Cambridge
\bibitem[\protect\citeauthoryear{Larson}{1974}]{larson74} Larson, R.B. 1974, MNRAS, 169, 229
\bibitem[\protect\citeauthoryear{Lea \& de Young}{1976}]{lea76}, Lea, S. M. \& de Young, D. S., 1976, ApJ, 210, 647
\bibitem[\protect\citeauthoryear{Li et al.}{2007}]{li07} Li, Z., Wang, Q. D., \& Hameed, S. 2007, MNRAS, 376, 960
\bibitem[\protect\citeauthoryear{Li \& Wang}{2007}]{li07_2} Li, Z., \& Wang, Q. D. 2007, ApJ, 668, L39
\bibitem[\protect\citeauthoryear{Li et al.}{2011}]{li11} Li, Z., Jones C., Froman, W.R., Kraft R.P., Lal, D.V., Stefano, R.D., Spitler L.R., Tang S., Wang Q.D., Gilfanov, M., \& Revnivtsev M. 2011, ApJ, 730, 84
\bibitem[\protect\citeauthoryear{Lynds \& Sandage}{1963}]{lynds63} Lynds C. R., Sandage A. R., 1963, ApJ, 137, 1005
\bibitem[\protect\citeauthoryear{Macci\`o et al. }{2008}]{maccio08} Macci\`o, A.V., Dutton, A.A., \& van den Bosch, F.C. 2008, MNRAS, 391, 1940 
%%\bibitem[\protect\citeauthoryear{McConnell et al.}{2011}]{mcconnell11}  McConnell, N.J., Ma, C-P., Gebhardt, K., Wright, S.A., Murphy, J.D., Lauer, T.R., Graham, J.R. \& Richstone, D.O., 2011, Nature, arXiv:1112.1078
\bibitem[\protect\citeauthoryear{Mannucci et al.}{2005}]{mannucci05} Mannucci, F., Della Valle, M., Panagia, N., Cappelarro, E., Cresci, G., Maiolino, R., R., Petrosian, A., \& Turatto, M. 2005, A\&A, 433, 807
\bibitem[\protect\citeauthoryear{Marconi \& Hunt}{2003}]{marconi03} Marconi, A., \& Hunt, L. K. 2003, ApJ, 589, L21
\bibitem[\protect\citeauthoryear{Marinacci et al.}{2010}]{marinacci10} Marinacci F., Binney J., Fraternali F., Nipoti C., Ciotti L., Londrillo P., 2010, MNRAS, 404, 1464
\bibitem[\protect\citeauthoryear{Marinacci et al.}{2011}]{marinacci11} Marinacci F., Fraternali F., Nipoti C., Binney J., Ciotti L., Londrillo P., 2011, MNRAS, 415, 1534
\bibitem[\protect\citeauthoryear{Martin}{2006}]{martin06} Martin, C. L. 2006, ApJ, 647, 222
\bibitem[\protect\citeauthoryear{Mathews \& Baker}{1971}]{mathews71} Mathews, W. G., \& Baker, J. C. 1971, ApJ, 170, 241
\bibitem[\protect\citeauthoryear{Moore et al.}{1999}]{moore99} Moore, B., Quinn, T., Governato, F., Stadel, J., \& Lake, G. 1999, MNRAS, 310, 1147
\bibitem[\protect\citeauthoryear{Mori et al.}{1997}]{mori97} Mori, M., Yoshii, Y., Tsujimoto, T., \& Nomoto, K. 1997, ApJ, 478, L21
\bibitem[\protect\citeauthoryear{Mori, Yoshii \& Nomoto}{1999}]{mori99} Mori, M., Yoshii Y., \& Nomoto, K. 1999, ApJ, 511, 585
\bibitem[\protect\citeauthoryear{Mori, Ferrara \& Madau}{2002}]{mori02} Mori, M., Ferrara, A., \& Madau, P. 2002, ApJ, 571, 40
\bibitem[\protect\citeauthoryear{Mori \& Umemura}{2006}]{mori06} Mori, M., \& Umemura, M., 2006, Nature, 440, 644
\bibitem[\protect\citeauthoryear{Moster et al.}{2010}]{moster10} Moster, B.P., Somerville, R.S., Maulbetsch, C.,  et al. 2010, ApJ, 710, 903
\bibitem[\protect\citeauthoryear{Navarro et al.}{1996}]{navarro96} Navarro, J.F., Frenk, C.S., \& White, S.D.M. 1996, Apj, 462, 563
\bibitem[\protect\citeauthoryear{Neugebauer \& Snyder}{1962}]{neugebauer62} Neugebauer, M., \& Snyder, C. W. 1962, Science, 138, 1095
\bibitem[\protect\citeauthoryear{Ogiya \& Mori}{2011}]{ogiya11} Ogiya, G., \& Mori, M. 2011, ApJ, 736, 2
\bibitem[\protect\citeauthoryear{Ogiya et al.}{2014}]{ogiya14_2} Ogiya, G., Mori, M., Ishiyama, T., \& Burkert, A. 2014, MNRAS, 440, 71
\bibitem[\protect\citeauthoryear{Ogiya \& Mori}{2014}]{ogiya14} Ogiya, G., \& Mori, M. 2014, ApJ, 793, 46
\bibitem[\protect\citeauthoryear{Ogiya \& Burkert}{2015}]{ogiya15} Ogiya, G., \& Burkert, A. 2015, MNRAS, 446, 2363
\bibitem[\protect\citeauthoryear{Oppenheimer \& Dav\'{e}}{2006}]{oppenheimer06} Oppenheimer, B.D., \& Dav\'{e}, R. 2006, MNRAS, 373, 1265 
\bibitem[\protect\citeauthoryear{Osterbrock}{1960}]{osterbrock60} Osterbrock, D.E., 1960, ApJ, 132, 325
\bibitem[\protect\citeauthoryear{Parker}{1958}]{parker58} Parker, E.N, 1958, ApJ, 128, 664
\bibitem[\protect\citeauthoryear{Parker}{1965}]{parker65} Parker, E.N, 1965, Space Sci. Rev., 4, 666
\bibitem[\protect\citeauthoryear{Pellegrini et al.}{2012}]{pellegrini12} Pellegrini, S., Wang, J., Fabbiano, G., Kim, D-W., Brassington, N.J., Gallagher, J.S., Trinchieri, G., \& Zezas, A. 2012, ApJ, 758, 94
\bibitem[\protect\citeauthoryear{Pettini et al.}{2001}]{pettini01} Pettini, M., Shaley, A.E., Steidel, C.C., Cuby, J.-G., Dickinson, M., Moorwood, A.F.M., Adelberger, K.L., \& Giavalisco, M. 2001, ApJ, 554, 981
\bibitem[\protect\citeauthoryear{Prada et al.}{2012}]{prada12} Prada, F., Klypin, A.A., Cuesta, A.J., Betancort-Rijo, J.E. \& Primack, J., 2012, MNRAS, 423, 3018
\bibitem[\protect\citeauthoryear{Ptuskin et al.}{1997}]{ptuskin97} Ptuskin V. S., Voelk H. J., Zirakashvili V. N., Breitschwerdt D., 1997, A\&A, 321, 434
\bibitem[\protect\citeauthoryear{Puchwein \& Springel}{2012}]{puchwein12} Puchwein, E., \& Springel, V. 2013, MNRAS, 428, 2966
\bibitem[\protect\citeauthoryear{Reddy et al.}{2006}]{reddy06} Reddy, B. E., Lambert, D.L., \& Allende, Pietro, C. 2006, MNRAS, 367, 1329
\bibitem[\protect\citeauthoryear{Sakamoto, Chiba \& Beers}{2003}]{sakamoto03} Sakamoto, T., Chiba, M. \& Beers, T. C. 2003, A\&A, 397, 911
\bibitem[\protect\citeauthoryear{Shapiro \& Field}{1976}]{shapiro76} Shapiro P. R, \&Field G.B. 1976, ApJ, 205, 762
\bibitem[\protect\citeauthoryear{Shapley et al.}{2003}]{shapley03} Shapley, A.E., Steidel, C.C., Pettini, M., \& Adelberger, K.L. 2003, ApJ, 588, 65
\bibitem[\protect\citeauthoryear{Sharma \& Nath}{2012}]{sharma12} Sharma, M., \& Nath, B.B. 2012, ApJ, 750, 55
\bibitem[\protect\citeauthoryear{Sharma \& Nath}{2013}]{sharma13} Sharma, M., \& Nath, B.B. 2013, ApJ, 763, 16
\bibitem[\protect\citeauthoryear{Sharma et al.}{2014}]{sharma14} Sharma, M. Nath, B. B.,  Chattopadhyay, I., \& Shchekinov, Y. 2014, MNRAS, 441, 431
\bibitem[\protect\citeauthoryear{Silk \& Rees}{1998}]{silk98} Silk, J., \& Rees, M.J. 1998, A\&A, 331, L1
\bibitem[\protect\citeauthoryear{Somerville et al.}{2008}]{somerville08} Somerville, R.S., Hopkins, P.F., Cox, T.J., Robertson, B.E., \& Hernquist, L. 2008, MNRAS, 391,481
\bibitem[\protect\citeauthoryear{Songaila}{1997}]{songaila97} Songaila, A., 1997, ApJ, 490, L1
\bibitem[\protect\citeauthoryear{Spitzer \& Baade}{1951}]{spitzer51} Spitzer, L. Jr. \& Baade, W., 1951, ApJ, 113, 413
\bibitem[\protect\citeauthoryear{Strickland}{2002}]{strickland02} Strickland, D. 2002, Astronomical Society of the Pacific Conference Series, 253, 387
\bibitem[\protect\citeauthoryear{Strickland et al.}{2004}]{strickland04} Strickland, D.K., Heckman, T.M., Colbert, E.J.M., Hoopes, C.G., \& Weaver, K.A. 2004, ApJS, 151, 193
%%\bibitem[\protect\citeauthoryear{Tremaine et al.}{2002}]{tremaine02} Tremaine, S., Gebhardt, K., Bender, R., et al. 2002, ApJ, 574, 740
\bibitem[\protect\citeauthoryear{Tsuchiya et al.}{2013}]{tsuchiya13} Tsuchiya, S., Mori, M., \& Nitta, S. 2013, MNRAS, 432, 2837
\bibitem[\protect\citeauthoryear{Tsujimoto}{2007}]{tsujimoto07} Tsujimoto, T. 2007, ApJ, 665, 115
\bibitem[\protect\citeauthoryear{Tsuru et al.}{2007}]{tsuru07} Tsuru, T.G.,  et al., 2007, PASJ, 59, S269
\bibitem[\protect\citeauthoryear{Uhlig et al.}{2012}]{uhlig12} Uhlig, M, Pfrommer, C., Sharma, M., Nath B. B., En$\beta$lin, T.A., \& Springel V. 2012, MNRAS, 423, 2374
\bibitem[\protect\citeauthoryear{de Vaucouleurs}{1948}]{devaucouleurs48} de Vaucouleurs, G. 1948, Ann. d'Ap., 11, 247
\bibitem[\protect\citeauthoryear{van Woerden \& Wakker}{2004}]{woerden04} van Woerden H., \& Wakker B. P. 2004, Astrophysics and Space Science Library, 312, 195
\bibitem[\protect\citeauthoryear{Veilleux et al.}{2005}]{veilleux05} Veilleux, S., Cecil, G., \& Bland-Hawthorn, J. 2005, ARA\&A, 43, 769
\bibitem[\protect\citeauthoryear{Wang}{1995}]{wang95} Wang, B. 1995, ApJ, 444, 590
\bibitem[\protect\citeauthoryear{Weiner et al.}{2009}]{weiner09} Wriner, B.J., et al. 2009, ApJ, 692, 187
\bibitem[\protect\citeauthoryear{White \& Rees}{1978}]{white78} White, S.D.M. \& Rees, M.J. 1978, MNRAS, 183, 341
\bibitem[\protect\citeauthoryear{White \& Frenk}{1991}]{white91} White, S.D.M. \& Frenk, C.S. 1991, ApJ. 379, 52
%%\bibitem[\protect\citeauthoryear{Xiao et al.}{2011}]{xiao11} Xiao, T., Barth, A., Greene, J.E., Ho, L.C., Bentz, M.C., Ludwig, R.R. \& Jiang, Y., 2011, ApJ, 739, 28 
\bibitem[\protect\citeauthoryear{Zasov}{1975}]{zasov75} Zasov, A. V., 1975, SvA, 18, 426
\bibitem[\protect\citeauthoryear{Zirakashvili et al.}{1996}]{zirakashvili96} Zirakashvili V. N., Breitschwerdt D., Ptuskin V. S., Voelk H. J., 1996, A\&A, 311, 113

\end{thebibliography}
\end{document}